\documentclass[a4paper,11pt]{article}
\pdfoutput=1 
\usepackage{graphicx}
\usepackage{dcolumn}
\usepackage{bm}
\usepackage{amssymb}
\usepackage{graphicx} 
\usepackage{amsbsy}
\usepackage{amsmath} 
\usepackage{slashed}
\usepackage{amsfonts}
\usepackage{comment}
\usepackage{braket}
\usepackage{tensor}
\usepackage[scriptsize]{subfigure}
\usepackage{float}
\usepackage{enumerate}
\usepackage{multirow}
\usepackage{yfonts}
\usepackage{breqn}
\usepackage{cancel}
\usepackage{pifont}

\usepackage{tikz}
\usetikzlibrary{shapes,arrows,positioning,automata,backgrounds,calc,er,patterns}
\usepackage[compat=1.1.0]{tikz-feynman}
\usetikzlibrary{positioning}
\usepackage{tikz-feynman}
\usepackage{makecell}

\usepackage{jheppub}

\usepackage[utf8]{inputenc}

\usepackage{listings}
\usepackage{color}

\definecolor{mygreen}{rgb}{0,0.6,0}
\definecolor{mygray}{rgb}{0.5,0.5,0.5}
\definecolor{mymauve}{rgb}{0.58,0,0.82}

\definecolor{darkWhite}{rgb}{0.94,0.94,0.94}

\newcommand{\dd} {\mathrm{d}}
\newcommand{\tr} {\mathrm{tr}}
\newcommand{\Tr} {\mathrm{Tr}}
\newcommand{\dc} {\mathcal{D}}

\newcommand{\g}{\sqrt{ g }}

\newcommand{\sD}{\slashed{D}}

\newcommand{\vev}[1]{\langle #1 \rangle}

\newcommand{\eps}{\epsilon}

\newcommand{\sig}{ \sigma}

\newcommand{\Acal}{{\mathcal{A}}}

\usepackage{dsfont}

\usepackage{hyperref}

\usepackage{color}
\definecolor{verdes}{cmyk}{0.92,0,0.59,0.4}

\definecolor{Grn}{rgb}{0.1,0.5,0.2}
\definecolor{Blu}{rgb}{0.,0.,0.1}
\definecolor{Red}{rgb}{0.7,0.1,0.1}
\definecolor{SE}{rgb}{0.5,0,0.4}
\definecolor{Tur}{rgb}{0,0.75,0.65}
\definecolor{Ceru}{RGB}{107,188,218}

\newcommand{\Red}[1]{{\color{Red}{#1}}}

\usepackage{hyperref} 
\hypersetup{
     colorlinks = true,
     linkcolor = black,
     citecolor = blue,
     anchorcolor = blue,
     filecolor = blue,
     urlcolor = blue,
     bookmarks=true,
     linktocpage =true,
     }

\renewcommand{\vec}[1]{\mathbf{#1}}

\newcommand{\dime}{d}

\newcommand{\Det}{\operatorname{Det}}

\usepackage{xparse}

\NewDocumentCommand{\dkpara}{O{k} O{\dime}}{\frac{{\rm d} {#1}^{\parallel}}{(2\pi)^{#2}}}
\NewDocumentCommand{\dkd}{O{k} O{\dime}}{ \frac{{\rm d}^{#2} {#1}}{(2\pi)^{#2}}}
\NewDocumentCommand{\dxd}{O{x} O{\dime}}{ {\rm d}^{#2} {#1} }
\NewDocumentCommand{\ad}{O{} O{}}{ \bigl \langle {#1}\bigr\rangle_{{\rm ad}{#2}} }

\begin{document}

\title{\boldmath Anomalous Transport from Effective Field Theory}

\author [1]{Rémy Larue} 
\author [2]{Amaury Marchon}
\author [2]{J\'er\'emie Quevillon} 
\author [2,3]{and Diego Saviot}

\affiliation[1]{School of Physical Science and Technology, ShanghaiTech University,
393 Middle Huaxia Road, Shanghai 201210, China}

\affiliation[2]{Laboratoire d’Annecy de Physique Th\'eorique,
CNRS – USMB, BP 110 Annecy-le-Vieux, 74941 Annecy, France}

\affiliation[3]{Laboratoire de Physique Subatomique et de Cosmologie, Universit\'e Grenoble-Alpes, CNRS/IN2P3, Grenoble INP, 38000 Grenoble, France}

\emailAdd{ryrlarue@shanghaitech.edu.cn}
\emailAdd{amaury.marchon@lapth.cnrs.fr}
\emailAdd{jeremie.quevillon@lapth.cnrs.fr}
\emailAdd{diego.saviot@lapth.cnrs.fr}

\abstract{
A systematic study of chiral effects is presented using an Effective Field Theory framework.
By integrating out a massive Dirac fermion at finite temperature in presence of vector and axial background fields, the currents and their anomalies are computed from the path-integral.
Chiral effects previously considered separately naturally arise in a unified computation, including new mass corrections.
The link between each anomalous transport effect and the anomalies is clearly established, beyond the identification of their coefficients.  In particular, we can appreciate how these effects are sourced by the anomalous nature of the theory even in configurations where the anomaly  itself vanishes.
The consistent and covariant anomalies are both encapsulated in master formulae for the currents which result from a careful treatment of the regularisation. It is finally found that, at finite temperature, the physical currents cannot be inferred simply from the Chern-Simons terms whose divergence reproduce the chiral anomalies.
}

\maketitle

\newpage
\section{Introduction}

The symmetries of a system play a fundamental role in the emergence of its low-energy hydrodynamic modes. The presence of a quantum anomaly, that is to say the violation of such a symmetry by quantum fluctuations, can therefore leave a strong imprint on the hydrodynamic transport effects. 
The chiral anomaly is known to have significant impact on transport phenomena, which are relevant to many physical systems such as quark gluon plasma or Weyl and Dirac semimetals in condensed matter. Surprisingly, its implications in astrophysical and cosmological contexts, where the relevant physical ingredients naturally coexist, remain relatively unexplored.

The theoretical development of these chiral effects started with \cite{Vilenkin:1978_blackholes,Vilenkin_1978_vortical_current,Vilenkin:1979}. The importance of the Chiral Magnetic Effect (CME) was later understood in~\cite{Fukushima:2008xe}, as well as the Chiral Separation Effect (CSE)~\cite{Metlitski:2005pr}, whereas general considerations on the impact of anomalies on the hydrodynamic construction was undertaken in~\cite{Son:2009tf}, including the relevance of the Chiral Vortical Effect (CVE).
They all appear as parity-violating responses of a system at finite temperature and density in presence of external fields, in particular electromagnetic fields.
More generally, external axial fields may also be considered. 
In condensed matter, they are known as pseudo-electromagnetic fields~\cite{Ilan:2019lqk} which appear in strained material as an intrinsic magnetisation and  affect the band structure of the material. In particular, they can contribute to the Anomalous Hall Effect (AHE)~\cite{Nagaosa:2009ycg}.

In the common hydrodynamic approach, the low-energy effective action is constructed out of all possible operators consistent with a set of symmetries, and organised order by order in derivatives of the hydrodynamic variables and external fields. The currents relevant for transport phenomena can then be extracted from this effective action. The strength of this approach is that it remains agnostic about the details of the microscopic theory, however the drawback is that the anomaly has to be manually inserted. While some transport coefficients can be constrained from general principles, eventually the specification of the microscopic theory is necessary to evaluate them exactly. This is achieved using diagrammatic methods derived from Kubo formulas targeting a specific operator~\cite{Kubo:1957mj,Kubo:1957wcy}. Provided one knows which operators arise from the anomaly, their coefficients can thus be obtained one-by-one. Nonetheless, it is in general not straightforward to identify which operators actually arise from an anomaly. For example, the link between the CVE and the chiral anomaly is highly non-trivial and has been the subject of many studies. In the hydrodynamic approach, its coefficient is not constrained enough to fix the temperature dependence. Indirect computations relying on holographic setups found a link with the axial-gravitational anomaly of a higher-dimensional space \cite{Amado:2011zx,Landsteiner:2011iq,Stone:2018zel,Landsteiner:2022wap}. These difficulties motivate a direct derivation of the effective action and the anomaly, as was done in some of the author's previous work~\cite{Larue:2025yar}.
Another difficulty when dealing with anomalies is the regularisation of the theory. This led to discrepancies and debates in the literature in particular concerning the absence of CME at equilibrium~\cite{Rubakov:2010qi,Yamamoto:2015fxa,Landsteiner:2016led,Zubkov:2016tcp,Zhang:2019yac, Brandt:2024wlw,Brandt:2024fpc}. It was noted in~\cite{Rubakov:2010qi,Landsteiner:2016led} that these discrepancies were tied to the so-called Bardeen-Zumino counterterm~\cite{Bardeen:1984pm} which relates covariant and consistent forms of the anomalies. This highlights that the Kubo formulae are not particularly well-suited to deal with these ambiguities and complicate the evaluation of the whole anomaly.
This further motivates a direct evaluation of the vector and axial currents as well as their associated anomalies with vector and axial backgrounds in a framework that allows for a careful treatment of the regularisation, which is the goal of this paper.

Here we focus only on continuous symmetries which can be treated perturbatively but hydrodynamics consequences of non-perturbative anomalies (for instance with discrete global symmetries) have been studied in \cite{Davighi:2023mzg,Davighi:2024qmh}. We rely on the one-loop effective action expressed as a path integral over the fermion field to derive the vector and axial currents and their associated anomalies. The temperature is introduced as an imaginary-time in the action following Matsubara's formalism \cite{Matsubara:1955ws}, turning the effective action into the thermodynamic partition function. It is then expanded in powers of temperature using a Covariant Derivative Expansion (CDE)~\cite{Gaillard:1985uh,Cheyette:1985ue,Larue:2025yar} which was adapted to finite temperature and densities on a curved spacetime in Ref.~\cite{Larue:2025yar}. This formalism also naturally allows us to obtain the corrections due to a fermion mass.
Similar methods can be found in \cite{Fuentes-Martin:2026bhr} with the notion of dimensional reduction particularly relevant to the study of cosmological phase transitions.

Technically, our main results are master formulae for the currents and anomalies at finite mass and temperature, where the freedom of regularisation is encoded in a set of three free parameters.  This includes new mass corrections to the transport effects. Conceptually, our analysis of the different $m/T$ regimes unravels the limitations of identifying the transport currents with the Chern-Simons currents, as is often done in the literature. In particular, we provide the first direct evaluation and confirmation of some transport effects which were postulated
on the basis of a Chern-Simons currents analysis \cite{Zyuzin:2012tv,Huang:2017rpa}, namely the Chiral pseudo-Magnetic Effect (CpME), Chiral pseudo-Separation Effect (CpSE) and Anomalous Axial-Hall Effect (AAHE). However, as we will show in Sec.~(4.4), this analysis fails at finite temperature. Our effective theory construction enables this systematic and self-sufficient analysis of all transport phenomena in hydrodynamics related to chiral anomalies.

In Sec.~\ref{sec:Partition}, we will introduce our framework to compute the vector an axial currents and anomalies from an imaginary-time path integral, and discuss the consistent and covariant regularisations.
The computations of the consistent vector and axial currents and anomalies will be carried out in Sec.~\ref{sec:Consistent}, first in the massless case, then with mass corrections.
In Sec.~\ref{sec:EliasAn}, we derive master formulae for the currents and anomalies in a regularisation independent manner, with an application to the covariant case. The consistent and covariant expressions are summarised in Tables~\ref{tab:vector} and \ref{tab:axial}. We then analyse our results in different regimes of the ratio $m/T$ and discuss their connection to the Chern-Simons currents.
The conclusions are presented in Sec.~\ref{sec:conclusion}.

\section{Partition function and anomaly in chiral background}\label{sec:Partition}

The partition function at local thermodynamic equilibrium and in flat spacetime with a fluid at rest is obtained by maximising the Gibbs entropy and reads
\begin{equation}
Z=\Tr\, e^{-\int_{\Sigma(t)}\dd^3x\,\beta(\hat H-\mu_a \hat j^t_a)}\;,
\label{eq:Ztr}
\end{equation}
where $\hat H$ is the Hamiltonian, $\beta$ the inverse temperature, $\hat j_a^\mu$ are a collection of currents with their associated chemical potentials $\mu_a$, and $\Sigma(t)$ is a time-slice.

We are interested in describing a fermion fluid at rest 
coupled to abelian vector and axial backgrounds. Then, using standard path integral techniques, the partition function~\eqref{eq:Ztr} may be recast as~\cite{Matsubara:1955ws,Bellac:2011kqa,Hongo:2016mqm,Hongo:2019rbd}
\begin{alignat}{4}   &Z&\;=\;&\int_{b.c}\dc\bar\psi\dc\psi e^{-\int_0^{\beta}\dd t\int\dd^3x\,\mathcal{L}}\,, \quad&\; &\;\mathcal L&\;=\;&\bar\psi\left(i\slashed D-m\right)\psi,\nonumber\\
&\slashed{D}&\;=\;&\gamma^{\mu}D_{\mu} \,,&\;&D_\mu&\;=\;&\partial_\mu+i V_\mu+i\gamma_5 A_\mu\;,\label{eq:Zpi}\\
&V_\mu&\;=\;& (\mu_V,\vec V) \,,\quad &\;&A_\mu&\;=\;&(\mu_A, \vec A)\;,\nonumber
\end{alignat}
where the boundary conditions are antiperiodic in imaginary time with period $\beta$, i.e., $b.c = \{ \psi(i\beta,\vec{x}) = -\psi(0,\vec{x}),\;\bar\psi(i\beta,\vec{x}) = -\bar\psi(0,\vec{x}) \}$.
Note that in thermodynamic equilibrium, the chemical potentials are indistinguishable from the time-component of their associated external field.~\footnote{Note that trying to absorb a chemical potential by a gauge transformation will simply shift it to the boundary conditions~\cite{Larue:2025yar}}
We also stress that the background fields and thermodynamic quantities are independent of the imaginary-time, only the path integral variables $\psi$ and $\bar\psi$ depend on it. In practice, this means that the imaginary-time formalism yields no time-derivatives in the expectation values (EVs), and therefore our computations are valid at local equilibrium in the adiabatic regime where the (real)-time-derivatives are negligible. We will thus only compute the spatial parts of the vector and axial currents as well as their spatial-divergence, which are defined in the following section.

\subsection{Vector current}
The vector current $j_V=\bar\psi\gamma^\mu\psi$ is associated with a $U(1)_V$ symmetry under which the path integral variables transform as
\begin{align}\label{eq:U1V}
    \psi'=e^{i\theta}\psi \,,\quad \bar\psi'=e^{-i\theta} \bar\psi \,,
\end{align}
with local parameter $\theta(\vec{x})$.

The partition function is invariant under this relabeling of the path integral variables, and after performing the functional integration we may write
\begin{equation}
Z=\Det_\beta[i\sD-m]=J_V[\theta]\Det_\beta[i\sD-m-\slashed\partial\theta]\;,\label{eq:JacDet}
\end{equation}
where the subscript $\beta$ indicates the $\beta$-anticyclicity of the functional integration. The Jacobian $J_V[\theta]$ arises due to the transformation of the measure and, if different from one, indicates the presence of a quantum anomaly $\mathcal{A}_V$~\cite{Fujikawa:2004cx}. It is thus expressed as a ratio of determinants~\cite{Filoche:2022dxl} and can be recast as
\begin{equation}\label{eq:JVtr}
\int_0^\beta\dd t\,\mathcal{A}_V(\textbf{x})=-\frac{\delta}{\delta\theta}\log J_V[\theta]=-\frac{\delta}{\delta\theta}\Tr_\beta \left[(\slashed\partial \theta) \frac{1}{i\slashed D - m} \right]\;.
\end{equation}
Note that the integrals over imaginary-time are trivial in these expressions and lead to a simple factor $\beta$.
Eq.~\eqref{eq:JVtr} is simply the expression of the anomalous Ward identity with time-independent background fields
\begin{align}
\partial_i\vev{j^i_V}=\mathcal{A}_V\;,
\end{align}
where $\vev{\cdot}$ denotes EV at finite temperature $T=1/\beta$ defined as
\begin{equation}\label{eq:definition_of_ev}
\vev{\mathcal{O}}\equiv\frac{1}{Z}\mathrm{Tr}\left[ \hat{\mathcal{O}}\,e^{-\int d^3 x\,\beta (\hat{H}-\mu_a\hat{j}_a^t)}\right]= \frac{1}{Z}\int_{b.c}\dc\bar\psi\dc\psi \,\mathcal{O}\, e^{-\int_0^\beta\int\dd^3x\,\mathcal{L}}\;.
\end{equation}
The EV of the current is identified before integrating by parts the derivative on $\theta$ in~\eqref{eq:JVtr}, or alternatively as
\begin{equation}
\vev{j^i_V}=\frac{\delta}{\delta\theta}\Tr_\beta\left[\theta\,\gamma^i \frac{1}{i\slashed D - m}\right]\;.
\end{equation}

\subsection{Axial current}\label{sec:introAxial}

Likewise, we obtain the EV of the axial current $j^\mu_A=\bar\psi\gamma^\mu\gamma_5\psi$ associated with the $U(1)_A$ axial symmetry
\begin{align}\label{eq:U1A}
    \psi'=e^{i\theta\gamma_5}\psi \,,\quad \bar\psi'= \bar\psi\,e^{i\theta\gamma_5} \,,
\end{align}
with local parameter $\theta(\textbf{x})$.

The crucial difference with the vector symmetry is that it is classically broken by the mass of the fermions. The anomaly breaking is thus given by the difference between the total and the classical breakings
\begin{align}\label{eq:defJAm}
    \int_0^\beta\dd t\,\mathcal{A}_A(\textbf{x})=-\frac{\delta}{\delta\theta}
    \log J_A[\theta]
    &=-\frac{\delta}{\delta\theta}\Tr_\beta \left[((\slashed\partial \theta)\gamma_5+2im\theta\gamma_5) \frac{1}{i\slashed D -m} \right] \,,
\end{align}
which is simply the expression of the Ward identity with time-independent background fields
\begin{equation}\label{eq:WardA}
    \partial_i\vev{j^i_A}=\mathcal{A}_A+2im\vev{\bar\psi\gamma_5\psi}\;,
\end{equation}
whereas the current itself is given by
\begin{align}\label{eq:jAeval}
    \langle j^i_A\rangle=\frac{\delta}{\delta\theta}\Tr_\beta \left[ \theta\,\gamma^i\gamma_5 \frac{1}{i\slashed D -m} \right] \,.
\end{align}
Note that the axial field may be viewed as non-gauged~\cite{Ilan:2019lqk}, in which case the axial symmetry breaking from the mass is not an issue. 
Alternatively, it may still be viewed as being associated with a gauge symmetry if one considers that the mass arises from a Spontaneous Symmetry Breaking such that the axial symmetry is classically restored by the introduction of the pseudo-Goldstone bosons. In that case, one may consider that we work in the unitary gauge where they do not explicitly appear.

\subsection{Ambiguities in the currents}\label{sec:ambiguities}

In the absence of a dynamical axial gauge field, the vector gauge symmetry can be preserved while only the global axial symmetry is broken. The theory thus remains consistent at the quantum level. 
However, when the axial symmetry is gauged, both gauge symmetries cannot be preserved at the quantum level. This leads to an ambiguity in the definition of the vector and axial currents corresponding to the freedom to break either gauge symmetry ~\cite{Bardeen:1969md,Bardeen:1984pm,Bertlmann:1996xk,Fujikawa:2004cx,Filoche:2022dxl}. 
In the literature, two forms of the anomalies are usually considered:

\begin{itemize}
    \item One may choose to preserve the vector symmetry. This leads to the so-called consistent form of the anomalies that respect the Wess-Zumino consistency conditions~\cite{Wess:1971yu}. From a diagrammatic point of view it amounts to enforcing the Bose symmetry of the diagrams. In the vacuum and in an abelian theory, the consistent anomaly reads 
    \begin{equation}\label{eq:Acons}
        \partial_\mu {\vev{j_{V,\mathrm{cons}}^\mu}}_{\mathrm{vac}}=0\;,\quad\quad \partial_\mu {\vev{j_{A,\mathrm{cons}}^\mu}}_{\mathrm{vac}}=\frac{1}{8\pi^2}\left(F_V \tilde F_V+\frac{1}{3}F_A \tilde F_A\right)\;.
    \end{equation}
    where $F_{V,\mu\nu}=\partial_\mu V_\nu-\partial_\nu V_\mu$ and $\tilde F_V^{\mu\nu}=\frac{1}{2}\epsilon^{\mu\nu\rho\sigma}F_{V,\rho\sigma}$, and likewise for $F_A$ and $\tilde F_A$. Note that although the consistent anomaly is axial-gauge invariant in an abelian theory, the consistent vector and axial currents are not, as we will see later on.~\footnote{\label{footnote1}In fact, in the abelian case, both the so-called covariant and consistent anomalies are gauge invariant and Wess-Zumino consistent (they are related by a local polynomial counterterm $\epsilon^{\mu\nu\rho\sigma}A_\mu V_\nu F^V_{\rho\sigma}$ in the action). However, only the consistent anomaly respects Bose symmetry.}
    \item On the other hand, one may choose to enforce that the currents have a covariant form, thereby computing the so-called covariant anomalies~\cite{Bardeen:1984pm,Fujikawa:2004cx}. This however results in both the axial and vector symmetries to be broken. In the vacuum and in an abelian theory, the covariant anomalies read
    \begin{equation}\label{eq:Acov}
        \partial_\mu {\vev{j_{V,\mathrm{cov}}^\mu}}_{\mathrm{vac}}=\frac{1}{4\pi^2}F_V \tilde F_A\;,\quad\quad \partial_\mu {\vev{j_{A,\mathrm{cov}}^\mu}}_{\mathrm{vac}}=\frac{1}{8\pi^2}(F_V \tilde F_V+F_A \tilde F_A)\;.   
    \end{equation}
\end{itemize} 
In practice, the ambiguities in the definition of the currents are tied to an ambiguity in the regularisation of the theory. In dimensional regularisation, this is due to the intrinsic 4-dimensional nature of $\gamma_5$. Different schemes exist in the literature which lead to different forms of the anomalies. One of the most widely-used is the Breitenlohner-Maison-t'Hooft-Veltman (BMHV) scheme~\cite{tHooft:1972tcz,Breitenlohner:1977hr}, which is known to yield the consistent anomaly~\cite{Bertlmann:1996xk, Novotny:1994yx, Horejsi:1988ei, Filoche:2022dxl}. In Sec.~\ref{sec:EliasAn}, we will also use the so-called Elias trick~\cite{Elias:1982ea,Filoche:2022dxl,Quevillon:2021sfz}, in which the ambiguity in the regularisation is handled by introducing free parameters and leads to our master formulae for the currents and anomalies. These give us the freedom to enforce either symmetry (preserve the vector or the axial), or enforce the covariance of the anomalies.

\subsection{Topological and non-topological anomalies}
\label{sec:topol}

The common lore concerning chiral anomalies is that they are independent from any intrinsic scale in the theory. In particular, it implies that at finite temperature and densities, they remain independent from both $T$ and $\mu_a$, i.e. only the vacuum contributes. As shown recently in Ref.~\cite{Larue:2025yar}, this statement has to be refined. 

Quantum anomalies can be separated into topological and non-topological pieces. The non-topological part can be subtracted by an adequate choice of regularisation and renormalisation, whereas the topological part cannot.~\footnote{To be more accurate, one should talk about non-trivial (instead of topological) vs trivial (instead of non-topological) anomalies. The distinction is irrelevant to the chiral anomalies, however, the Weyl anomaly has non-topological pieces which cannot be subtracted by renormalisation and are thus non-trivial.} The topological Pontryagin densities $F_{V/A}\tilde F_{V/A}$ are the topological parts of the abelian vector and axial anomalies, see Eqs.~\eqref{eq:Acons} and~\eqref{eq:Acov}. Importantly, since topological anomalies cannot be renormalised away, they are independent from the renormalisation scale $\lambda_{\mathrm{ren}}$. As a result, they are also independent from any intrinsic scale of the theory $\Lambda$, e.g. masses, since these run with the energy $\Lambda(\lambda_{\mathrm{ren}})$. However, nothing protects non-topological anomalies from intrinsic scales.

Although non-topological chiral anomalies are usually ignored, they can in fact be phenomenologically relevant. One such example occurs in the context of QCD where a non-topological $P$-even anomaly~\footnote{Note the abuse of language: by $P$-odd/even we refer to the structure of the operator, omitting the intrinsic parity of the fields.}, such as $\langle D j_A\rangle\supset (D_\mu A_\nu) A^\mu A^\nu$, is shown to be related to physical processes, such as the 4-pion interaction $\pi\pi\to\pi\pi$~\cite{Balog:1985ea}.

In this work we are interested in transport effects due to a vector and axial field together, at finite temperature, mass and chemical potentials. We will only consider parity-odd contributions to the vector and axial currents, and verify that the corresponding topological anomalies are independent from  mass and temperature.
We set aside all $P$-even, i.e. non-topological, contributions which will be studied in future work \cite{Larue:2026}.

\subsection{Global equilibrium with anomalies}

Although our computations are valid in the broader local equilibrium adiabatic regime, we may discuss the particular case of a global equilibrium, and how it is modified by the presence of anomalies.
We will consider the case of curved spacetime and non-zero fluid velocity for a more generic discussion. At a given time $t$, the information entropy is maximised by the generating functional
\begin{equation}
Z(t)=\Tr\,e^{\int_{\Sigma(t)}\dd\Sigma\,n_\mu(\hat T^{\mu\nu}\beta_\nu-\xi_a \hat j_a^{\mu})}\;,\label{eq:Zgeneric}
\end{equation}
where $\Sigma(t)$ is the time-slice at $t$ and $n$ its normal with $n^2=1$; $\beta^\mu=\beta u^\mu$ where $u$ with $u^2=1$ is the 4-velocity of the fluid with respect to the observer; $\beta$ is the (now position- and time-dependent) inverse  temperature and $\xi_a=\beta \mu_a$. Except for this Section, we consider in this paper the case where $n_\mu=(1,\vec 0)$, $u^\mu=(1,\vec 0)$ and the metric is Minkowski, such that Eq.~\eqref{eq:Zgeneric} becomes Eq.~\eqref{eq:Ztr}. We assume a symmetric Energy-Momentum Tensor (EMT) (i.e. no Lorentz anomaly).

At a later time $t+\Delta t$, in virtue of Stokes' theorem we have
\begin{align}
\begin{split}
&\int_{\Sigma(t+\Delta t)}\dd^3x\g \,n_\mu(\hat T^{\mu\nu}\beta_\nu-\xi_a\hat j^\mu_a)-\int_{\Sigma(t)}\dd^3x\g \,n_\mu(\hat T^{\mu\nu}\beta_\nu-\xi_a\hat j^\mu_a)\\
=&\int_{\Omega(t+\Delta t,t)}\dd^4x\g\left\{\hat T^{\mu\nu}D_\mu\beta_\nu-\hat j^\mu_a D_\mu \xi_a
-\xi_a D \hat{j}_a\right\} \label{eq:Omegatube}
\end{split}
\end{align}
where $\Omega(t+\Delta,t)$ is the 4-dimensional fluid tube enclosed by $\Sigma(t)$ and $\Sigma(t+\Delta t)$, and the flux of $\hat T^{\mu\nu}\beta_\nu-\xi_a \hat j_a^{\mu}$ is supposed to vanish on the timelike hypersurface joining them. 
To obtain~\eqref{eq:Omegatube}, we used the conservation equation of the EMT $D_\mu\hat T^{\mu\nu}=0$ (i.e. no diffeomorphism anomaly).~\footnote{\label{foot:muec}To match with the rest of the paper, we assumed $\partial_t V_\mu=0$ such that the electric field is the gradient of an electrostatic potential which we absorbed in the (electro-)chemical potential $\xi_a$, such that $D_\mu \xi_a$ is the total electric field. Equivalently, if $\xi_a$ were kept as the internal chemical potential only, the conservation equation would be $D_\mu\hat T^{\mu\nu}=F_a^{\mu\nu}\hat j_{a,\nu}$.}
The time variation of the partition function is then expressed as \cite{Hongo:2016mqm}
\begin{equation}
\frac{\dd}{\dd t} \log Z[t,V(x)]= \int\dd^3x\g\,\left(\vev{\hat T^{\mu\nu}}D_\mu\beta_\nu-\vev{\hat j^\mu_a}D_\mu\xi_a-\xi_aD_{\mu}\vev{\hat j^\mu_a}\right)\;.
\end{equation}
The global equilibrium conditions can be expressed as:
\begin{itemize}
    \item the 4-temperature is Killing $D_\mu\beta_\nu+D_\nu\beta_\mu=0$,
    \item $\vev{\hat j_a^\mu} D_\mu\xi_a+\xi_a D_{\mu}\vev{\hat j_a^{\mu}}=0$.
\end{itemize}
If $j^\mu_a$ are associated with non-anomalous symmetries then $D_{\mu}\hat j_a^{\mu}=0$, and we obtain the usual non-anomalous global equilibrium conditions: $\beta_\mu$ is Killing and the total electric fields vanish $D_\mu\xi_a=0$~\cite{Zubarev:1979afm,Weldon:1982aq,Weert_1982,Becattini:2012tc}.

For the case considered in this paper the second line  reads
\begin{equation}
\vev{j^\mu_V}\partial_\mu\mu_V+\mu_V\partial_\mu\vev{j_V^{\mu}}+\vev{j^\mu_A}\partial_\mu\mu_A+\mu_A\partial_\mu\vev{j_A^{\mu}} =0\;.\label{eq:anomalyGEQ}
\end{equation}
In general, EVs are non-local and can be expanded in an infinite series of polynomials in the background fields suppressed by a cut-off (temperature, chemical potential, mass,\,\dots), since they stem from the partition function which is itself non-local. This is the case of the currents $\vev{j_V}$ and $\vev{j_A}$. Crucially, anomalies are singular in that regard since they are local polynomials in the backgrounds fields. Therefore, there can be no systematic cancellation between the anomalies and the currents at all order in the expansion of the currents. Likewise, there can be no systematic cancellation between two infinite series unless they are proportional to each other, therefore $\vev{j^\mu_V}\partial_\mu\mu_V$ and $\vev{j^\mu_A}\partial_\mu\mu_A$ cannot cancel each other.

In a massless theory, the divergences of the currents $\partial_\mu\vev{j_V^{\mu}}$ and $\partial_\mu\vev{j_A^{\mu}}$ are purely anomalous. According to our discussion above, the global equilibrium condition Eq.~\eqref{eq:anomalyGEQ} can therefore only be met if the gradients of chemical potentials vanish identically
\begin{equation}\label{eq:globaleq}
    \partial \mu_V=\partial\mu_A=0\;.
\end{equation}
This leaves us with the following condition to satisfy
\begin{equation}\label{eq:GEQcondition}
\mu_V\partial\vev{j_V}+\mu_A\partial\vev{j_A}=\mu_V\mathcal{A}_V+\mu_A\mathcal{A}_A=0\;.
\end{equation}
However, anticipating on the results of this paper, the anomalies are always proportional to the electric fields $E=\partial\mu_V$ and $E_A=\partial\mu_A$, such that this condition is automatically met. \footnote{We postpone to future work~\cite{Larue:2026} the discussion concerning trivial anomalies and whether they could modify the gobal equilibrium conditions.
}

All in all, in the case of a purely anomalous breaking, the global equilibrium conditions are not modified and remain that the 4-temperature is Killing, and all gradients of chemical potentials must vanish (i.e the total electric fields must vanish).
Although the consistency of introducing chemical potentials $\mu_{V/A}$ is uncertain at local equilibrium due to the violation of the current conservation, they are well motivated at global equilibrium where the conservation of the currents is restored.~\footnote{It is also possible to modify the current associated with the axial chemical potential in the presence of the axial anomaly such that it is associated with a conserved current~\cite{Rubakov:2010qi}. According to our discussion this must not impact the global equilibrium.}

The massive case requires some care in discussing global equilibrium. According to Eq.~\eqref{eq:WardA}, the divergence of the axial current receives an explicit-breaking contribution $2im\vev{\bar\psi\gamma_5\psi}$. The pseudoscalar expectation value is, in general, a non-local functional of the background fields, which we only compute at leading order. There is, however, no reason to expect $\vev{\bar\psi\gamma_5\psi}$ to vanish to all orders in the absence of electric fields. If this expectation value is non-vanishing, the axial current is not conserved and, for $\mu_A\neq0$, the corresponding contribution enters the global-equilibrium condition in Eq.~\eqref{eq:GEQcondition}. Our imaginary-time formalism does not address the corresponding real-time relaxation dynamics. In particular, in the free theory considered here there is no interaction-induced relaxation mechanism, so the associated relaxation time is in practice infinite. Alternatively, let us propose two configurations where the global equilibrium is well-defined in our set-up. Firstly, one may consider that the mass arises from a spontaneous symmetry breaking mechanism such that the chiral symmetry is classically restored (see Sec.~\ref{sec:introAxial}). Secondly, one may simply not fix the axial charge for massive chiral fermions, i.e.~set $\mu_A=0$ while allowing $\vec A\neq0$, which may, for instance, be generated in Weyl semimetals. In this case, the non-conservation of the axial current entirely drops out of the global equilibrium condition which is hence well-defined. Most of the transport effects survive in this case. Anticipating the results of Tabs.~\eqref{tab:vector} and~\eqref{tab:axial}, we see that, for the consistent currents, only the CpSE (Chiral-pseudo Separation Effect) disappears, whereas the other transport effects acquire new mass corrections.

\section{Consistent currents and anomalies}\label{sec:Consistent}

In this Section, we explain our procedure to compute the EVs of the consistent currents and the anomalies, first in the massless case, and then with mass corrections. Note that even if the topological anomalies are mass- and temperature-independent, the corresponding currents are not in general.

\subsection{Massless case}

\subsubsection{Vector current}
The computational framework that will be used throughout this article is the Covariant Derivative Expansion (CDE)~\cite{Gaillard:1985uh, Cheyette:1985ue,Larue:2023uyv} which consists in evaluating the functional traces using a plane waves basis $e^{iq\cdot x}$. It was adapted to the imaginary-time formalism and curved spacetime in~\cite{Larue:2025yar}. The cyclicity of the time direction results in a discrete sum over the so-called (fermionic) Matsubara modes $\omega_n$.
Applying the CDE to expand the Jacobian of the $U(1)$ vector symmetry yields
\begin{align}
    \begin{split}\label{eq:JVCDE}
    \log J_V[\theta]
    &=-\int_0^{\beta}\!\!\!\!\dd t\int\!\!\dd^3x(\partial_\mu\theta)\,\frac{1}{\beta}\sum_{n\in\mathds Z}\int\frac{\dd^3 \vec q}{(2\pi)^3}\tr\,\gamma^\mu\sum_{k\in\mathds{N}}\left[\Delta i\sD\right]^k\Delta \,,
    \end{split}
\end{align}
where $\Delta$ is the free propagator
\begin{align}
    \Delta =\frac{\slashed q}{q^2} \,,\qquad q_\mu = (\omega_n,\vec q)\,,\qquad \omega_n = (2n+1)\pi T \,,\qquad \beta=\frac{1}{T}\,.
    \end{align}
The expansion \eqref{eq:JVCDE} is organised by mass dimension, and its truncation is possible under the assumption of large temperature. Note that this is only slightly different from the more common derivative expansion in hydrodynamics, or Ref.~\cite{Larue:2025yar}, in which the chemical potentials are included in the propagators hence treated on the same footing as the temperature~\footnote{A modification of the derivative expansion is argued to be necessary to avoid frame dependence of certain conductivities in \cite{Amoretti:2022vxq}, with the magnetic field taken to be of order zero too.}. Instead, we keep the chemical potentials as the time-component of the gauge fields, such that the expansion requires $\mu_a\ll T$. \footnote{In~\cite{Larue:2025yar}, the chemical potential is treated on the same footing as the temperature, and no corrections of order $\mu_a/T$ are found to the CSE. Likewise, we do not expect corrections to the other operators in the massless case.}

To more easily track the $\gamma_5$'s we define
\begin{align}
    D_\mu= D^V_\mu+i\gamma_5 A_\mu \,,\qquad D^V_\mu = \partial_\mu+iV_\mu \,.
\end{align}
The parity-odd contributions are the following
\begin{align}
    \begin{split}\label{eq:jVk3}
    \log J_V[\theta]=-\int_0^{\beta}\!\!\!\!\dd t\int\!\!\dd^3x(\partial_\mu\theta)\,\frac{1}{\beta}\sum_{n\in\mathds Z}\int\frac{\dd^3 \vec q}{(2\pi)^d}\tr\,\gamma^\mu (i\Delta \slashed{ A}\gamma_5 \Delta \slashed D^V\Delta \slashed D^V \Delta&\\
    + i\Delta \slashed D^V\Delta \slashed A\gamma_5\Delta \slashed D^V\Delta+i\Delta \slashed D^V\Delta \slashed D^V\Delta \slashed A\gamma_5-i(\Delta\slashed A\gamma_5)^3\Delta) &+\mathcal{O}(\beta^2) \,.
    \end{split}
\end{align}
Firstly, let us point out that because of the Levi-Civita tensor generated by the trace, the last term of~\eqref{eq:jVk3} vanishes for abelian gauge fields since $\epsilon^{ijk}A_iA_j A_k=\eps^{ijk}A_0 A_j A_k=0$.

We proceed with the computation in the following, deferring further details to App.~\ref{app:1}.
Anticipating the UV-divergences, we place ourselves in $d+1$ dimensions, where $d=3-\eps$ is the spatial dimension and $\epsilon$ will eventually be taken to zero. In conjunction, we will be using the BMHV-scheme for the computation of the Dirac algebra traces, as it is known to yield the consistent anomaly \cite{Bertlmann:1996xk, Novotny:1994yx, Horejsi:1988ei, Filoche:2022dxl}. Nevertheless, we will show in Section~\ref{sec:EliasAn} how to use a generic scheme that gives the freedom to preserve or violate either symmetry depending on the desired form of the anomaly.

The CDE procedure allows the separation of the operators from the momentum part. For example, the first term of \eqref{eq:jVk3} reads
\begin{align}\label{eq:jVex}
    \frac{1}{\beta}\sum_n \int\frac{\dd^d\vec q}{(2\pi)^d}\frac{q_\alpha q_\beta q_\gamma q_\delta}{(\omega_n^2+\vec q^2)^4}\tr \gamma^\mu \gamma^\alpha \gamma^\nu \gamma_5 \gamma^\beta \gamma^\rho \gamma^\gamma \gamma^\sigma \gamma^\delta (\partial_\mu \theta)A^\nu D^V_\rho V_\sigma \,.
\end{align}
We then separate the Matsubara frequencies $q_t=\omega_n$ from the spatial momenta $\mathbf{q}$ such that the expression can be expressed in terms of simple scalar master sum-integrals
\begin{align}\label{eq:defS}
    S_{a,b,c} = \frac{1}{\beta}\sum_{n\in\mathds Z}\int\frac{\dd^d\vec q}{(2\pi)^d}\frac{\vec q^{2a} \omega_n^{2b}}{(\omega_n^2+\vec q^2 + m^2)^c} \,,
\end{align}
with $m=0$ in this section. The master sum-integrals are computed analytically in
$d+1=4-\eps$ dimensions using a Mellin transformation in App.~\ref{app:1}.
Our computation then reduces to
\begin{align}
    \log J_V[\theta]
    =\;&4\int_0^{\beta}\!\!\!\!\dd t\int\!\!\dd^dx(\partial_i\theta)\eps^{ijk}
    \left[\; \left(3S_{0,0,2}-\frac{12}{d}S_{1,0,3}\right)A_0\partial_j V_k
    +\left(-S_{0,0,2}+4S_{0,1,3}\right)V_0\partial_j A_k\right.\nonumber\\
    &\hspace{1.3cm}\left. +\left(3S_{0,0,2}-4S_{0,1,3}-\frac{8}{d}S_{1,0,3}\right) A_j\partial_kV_0
    +\left(-S_{0,0,2}+\frac{4}{d}S_{1,0,3}\right)V_j\partial_k A_0 \right]_{m=0} \nonumber\\&\hspace{1.3cm} +\mathcal{O}(\beta^2) \,.\label{eq:Jvsumint}
\end{align}
In the following, we omit the $\mathcal{O}(\beta^2)$ for simplicity but it is understood.
The vector current can be read off~\eqref{eq:Jvsumint} in factor of $\partial_i\theta$, and upon using the expression of the master sum-integrals we obtain
\begin{align}
    \begin{split}\label{eq:jVm0cons}
    \langle\vec j_V\rangle = & 
    -\frac{1}{2\pi^2}\vec{A}\times \vec{E}
    +\frac{1}{2\pi^2}\mu_V\vec{B}_A \,,
    \end{split}
\end{align}
where $\vec{E}=\boldsymbol{\partial}\mu$ is the electric field and $\vec B_A = \boldsymbol{\partial}\times \vec A 
$ is the axial magnetic field. After integrating by part in \eqref{eq:Jvsumint}, i.e. taking the divergence of the vector current, the vector gauge anomaly is found to vanish identically
\begin{align}
    \begin{split}\label{eq:AVm0cons}
    \Acal_V[\theta]
    &=4 \left(2S_{0,0,2}-\frac{8}{d}S_{1,0,3}\right) \left(\vec E\cdot \vec B_A + \vec B\cdot \vec E_A\ \right) =0 \,.
    \end{split}
\end{align}
This is expected for the consistent anomaly~\eqref{eq:Acons} in which  the vector symmetry is preserved~\cite{Wess:1971yu,Bardeen:1984pm,Bertlmann:1996xk}.

Although only two type of operators occur in~\eqref{eq:jVm0cons}, on dimensional grounds all of the following $P$-odd operators could have appeared
\begin{alignat}{5}
    &\eps^{ijk} A_0\partial_jV_k &= &\mu_A \vec B^i &&\text{(Chiral Magnetic Effect - CME)}\,,\nonumber\\
    &\eps^{ijk} A_j\partial_kV_0 &= &(\vec A\times \vec E)^i &&\text{(Anomalous Hall Effect - AHE)}\,,&&\nonumber\\
    &\eps^{ijk}V_0\partial_j A_k &= &\mu_V \vec B_A^i &&\text{(Chiral pseudo-Magnetic Effect - CpME)}\,,
    \label{eq:VVAop}\\
    &\eps^{ijk}V_j\partial_k A_0 &= &(\vec V\times \vec E_A)^i \,,&&\nonumber\\
    &\eps^{ijk}V_0V_j A_k &= &\mu_V (\vec A\times \vec V)^i\,.&&\nonumber
\end{alignat}
The first three correspond to well-known transport effects: the CME~\cite{Fukushima:2008xe,Landsteiner:2011cp,Landsteiner:2016led}, the AHE \cite{Nagaosa:2009ycg,Zyuzin:2012tv} and the CpME \cite{Grushin:2016fgr} respectively. Importantly, to the best of our knowledge the CpME has only been postulated in the literature by identifying the transport currents with the Chern-Simons currents. As we will show in Sec.~\ref{sec:m/Tflow}, this identification fails at finite temperature. We thus provide for the first time a direct evaluation and confirmation of the CpME.
The last two terms violate the vector symmetry and are thus absent in the consistent currents. Note that the last operator was already absent from \eqref{eq:Jvsumint}, prior to the evaluation of the sum-integrals. Furthermore, it is of order zero in a derivative expansion whereas the others are of order one.~\footnote{It also has no $T=0$ equivalent whereas the others will be shown to be related to a decomposition of the Chern-Simons currents.}

The absence of the CME ($\vec{j}_V\propto \mu_A\vec{B}$) from the current is expected at equilibrium \cite{Rubakov:2010qi,Yamamoto:2015fxa,Landsteiner:2016led,Zubkov:2016tcp,Zhang:2019yac, Brandt:2024wlw,Brandt:2024fpc} due to the application of the Bloch theorem which forbids equilibrium current for a current associated to a gauge symmetry.

Note that the CpME and the AHE combine as a total divergence
\begin{align}
    \begin{split}
    \langle\vec j_V\rangle = & 
    -\frac{1}{2\pi^2}\vec{A}\times \vec{E}
    +\frac{1}{2\pi^2}\mu_V\vec{B}_A = \frac{1}{2\pi^2}\boldsymbol{\partial}\times(\mu_V \vec{A}) \,.
    \end{split}
\end{align}
This implies that the current integrated over the space $\vev{\vec{J}_V} = \int d^3x \vev{\vec{j}_V}$ can only be sourced by the boundary values of $\mu_V$ and $\vec A$. Nonetheless, $\vec A$ is not to be thought of as a gauge field since we will see shortly that the axial symmetry is broken. In fact, in condensed matter, it is not associated with a gauge symmetry~\cite{Ilan:2019lqk,Chernodub:2021nff}. For these reasons, it is a physical field that must vanish on the spatial boundary. The Bloch theorem
is therefore verified $\vev{\vec J_V}=0$ even in the presence of CpME.

On the other hand, having a non-vanishing CME without violating the Bloch theorem would require the presence of the $ \vec V\times \vec E_A $-term, such that $\vev{\vec j_V}\supset \boldsymbol\partial\times (\mu_A \vec V)$. This is however not allowed by vector gauge-invariance.

It is interesting to point out that, although the vector anomaly is vanishing, we have non-zero AHE and CpME in the vector current. They do depend on the regularisation in the same manner that anomaly does, that is to say they become finite without renormalisation (i.e they are of the form $\eps/\eps$ in dimensional regularisation).
More elements to improve our understanding will come in the following sections with the addition of a mass or with a more generic regularisation.

\subsubsection{Axial current}
Moving on to the axial current in the massless case, the CDE of the trace in \eqref{eq:jAeval} reads
\begin{align}\label{eq:CDEjA}
    \begin{split}
    \log J_A[\theta]&=-\int_0^{\beta}\!\!\!\!\dd t\int\!\!\dd^3x(\partial_\mu\theta)\,\frac{1}{\beta}\sum_{n\in\mathds Z}\int\frac{\dd^3 \vec q}{(2\pi)^3}\tr\,\gamma^\mu\gamma_5\sum_{k\in\mathds{N}}\left[\Delta i\sD\right]^k\Delta \,.
    \end{split}
\end{align}
We again consider only the leading order parity-odd contributions, i.e. at $k=3$ in the expansion.

We first focus on the terms that do not involve the axial field $A$. This corresponds to the ABJ anomaly which was computed by some of the authors in~\cite{Larue:2025yar} in curved spacetime. The relevant contribution from~\eqref{eq:CDEjA} is
\begin{align}
    \begin{split}
        &\tr\,\gamma^\mu\gamma_5 \Delta \slashed D^V \Delta \slashed D^V\Delta \slashed D^V \Delta \,.
    \end{split}
\end{align}
The different operators expected to arise from this term are
\begin{align}
\begin{split}
\eps^{ijk}V_0\partial_j V_k &= \mu_V \vec B^i \hspace{1.5cm}\text{(Chiral Separation Effect - CSE)}\,,\\
    \eps^{ijk} V_j\partial_k V_0 &= (\vec V\times \vec E)^i \,,
\end{split}
\end{align}
where in the first line we recognise the CSE~\cite{Metlitski:2005pr}, whereas the second term breaks the vector gauge symmetry. However, the second operator has a vanishing coefficient and we obtain
\begin{align}
    \begin{split}\label{eq:JaA=0}
    \left.\langle\vec j_A\rangle\right|_{A=0} = \frac{1}{2\pi^2}\mu_V \vec B \,.
    \end{split}
\end{align}
Again, this is expected from the use of the BMHV scheme which yields the consistent form of the anomalies where the vector symmetry is preserved.
At $m=0$ the axial anomaly is simply the divergence of the axial current and we find
\begin{align}
    \begin{split}\label{eq:AVcons}
    \left.\Acal_A[\theta]\right|_{A=0}=\frac{1}{2\pi^2}\vec E\cdot \vec B = \frac{1}{8\pi^2} F_V \tilde F_V \,.
    \end{split}
\end{align}

We now compute the contributions from the axial field
\begin{align}
    \begin{split}\label{eq:gamma5position}
        \tr\,\gamma^\mu\gamma_5(\Delta \slashed D^V \Delta& \slashed  A\gamma_5\Delta \slashed  A\gamma_5 \Delta 
        + \Delta \slashed  A\gamma_5 \Delta \slashed D^V \Delta \slashed  A\gamma_5 \Delta
        + \Delta \slashed  A\gamma_5\Delta \slashed  A\gamma_5 \Delta \slashed D^V \Delta) \\[4pt]
        =\frac{1}{(q^2)^4}q_\alpha q_\beta q_\gamma q_\delta \;&[\tr\, \gamma^{\mu}{}_5{}^{\alpha\nu\beta\rho}{}_5{}^{\gamma\sigma}{}_5{}^{\delta} ((\partial_\nu A_\rho) A_\sigma +  A_\rho (\partial_\nu  A_\sigma) +  V_\nu  A_\rho  A_\sigma) \\
        +&\tr\, \gamma^{\mu}{}_5{}^{\alpha\nu}{}_5{}^{\beta\rho\gamma\sigma}{}_5{}^{\delta} ( A_\nu (\partial_\rho  A_\sigma) +  A_\nu  V_\rho  A_\sigma) \\
        +&\tr\, \gamma^{\mu}{}_5{}^{\alpha\nu}{}_5{}^{\beta\rho}{}_5{}^{\gamma\sigma\delta} ( A_\nu  A_\rho  V_\sigma)] \,,
    \end{split}
\end{align}
where we used $\gamma^{ab\dots c}=\gamma^a\gamma^b\dots\gamma^c$ as a shorthand. The expected operators are listed below,
\begin{alignat}{4}
    &\eps^{ijk} A_0\partial_jA_k &= &\mu_A \vec B_A^i &&\text{(Chiral pseudo-Separation Effect - CpSE)} \,,\nonumber\\
    &\eps^{ijk} A_j\partial_kA_0 &= &(\vec A\times \vec E_A)^i \qquad&&\text{(Anomalous Axial-Hall Effect - AAHE)}\label{eq:JaOps}\,,&&\\
    &\eps^{ijk}A_0A_j V_k &= &\mu_A (\vec A\times \vec V)^i\,.&&\nonumber
\end{alignat}
The first two operators have been previously considered in~\cite{Huang:2017rpa} and labelled respectively Axial pseudo-Separation Effect (we call it CpSE in analogy with the CSE) and AAHE. As for the CpME, these transport effects were only postulated by identifying the transport currents with the Chern-Simons currents, although the identification fails at finite temperature (see Sec.~\ref{sec:m/Tflow}). To the best of our knowledge, this is the first direct evaluation and confirmation of these effects.
Including~\eqref{eq:JaA=0}, the axial current reads
\begin{align}
    \begin{split}
    \langle\vec j_A\rangle = & \frac{1}{2\pi^2}\mu_V \vec B+\frac{1}{3\pi^2}\mu_A \vec B_A -\frac{1}{6\pi^2}\vec A\times \vec E_A \,.
    \end{split}
\end{align}
Again, using the BMHV scheme, we notice the absence of the last operator of~\eqref{eq:JaOps} which breaks the vector gauge symmetry.

As for the axial anomaly, for $m=0$ it is given by the divergence of the axial current and we find
\begin{align}
    \begin{split}
    \Acal_A[\theta] = \frac{1}{2\pi^2}\left(\vec E\cdot\vec B+\frac{1}{3}\vec E_A\cdot \vec B_A\right) =  \frac{1}{8\pi^2} \left(F_V\tilde F_V+\frac{1}{3}F_A \tilde F_A\right) \,,
    \end{split}
\end{align}
where the relative $1/3$-factor is usual for the consistent form of the anomaly~\cite{Bardeen:1969md,Bertlmann:1996xk}.

\subsection{Massive case}
We now carry on the same computations but for a massive fermion. The propagator is now
\begin{equation}
\Delta =-\frac{-\slashed q +m}{q^2+m^2}\;.
\end{equation}
The effective action depends on two IR scales: $T$ and $m$. We are mostly interested in the case where the mass is small compared to the temperature $T\gg m$, such that the CDE will be expanded in powers of $\beta=1/T$.

In an effective action expanded in one parameter, the mass dimension of the effective operators is fixed by the order in the expansion, that is to say
\begin{equation}
\beta^{n}\vev{\mathcal{C}}=\sum_{i,j} c_{i,j}\beta^i\mathcal{O}_j^{(i)}\;,\quad [\mathcal{O}^{(i)}]=i\;,\quad [\mathcal{C}] =n\;,\quad [\beta]=-1\;,
\end{equation}
where $[\cdot]$ denotes the mass dimension, the $c_i$ are dimensionless coefficients, and $j$ runs over a basis of operators of a given mass dimension. On the other hand in an expansion with two dimensionful parameters, the structure is much more complex since operators of different mass dimensions can now contribute at the same order in $\beta$. In practice our expansion is organised as
\begin{equation}
\beta^{n}\vev{\mathcal{C}}=\sum_{i,j} c_{i,j}(\beta m)\;\beta^i \mathcal{O}_j^{(i)}\,, \quad c(\beta m) = \sum_k a_{k}\, (\beta m)^k\;,
\end{equation}
where the coefficients $c_{i,j}$ become functions of the new dimensionless parameter $\beta m$.
In the following, we compute the mass corrections in $(\beta m)^j$ to the transport coefficients, and verify that the anomalies are unaltered.

\subsubsection{Vector current}
The presence of $m$ in the numerator of the propagators leads to additional terms. Going back to \eqref{eq:jVk3} prior to the evaluation of the traces, the first term for example now reads
\begin{align}
    &\tr\,\gamma^\mu \Delta \slashed{ A}\gamma_5 \Delta \slashed D^V\Delta \slashed D^V \Delta \nonumber\\
    =&\frac{1}{(q^2+m^2)^4}\Big[\; m^4 \tr \gamma^{\mu\nu}{}_5{}^{\rho\sigma} + q_\alpha q_\beta q_\gamma q_\delta \tr \gamma^{\mu\alpha\nu}{}_5{}^{\beta\rho\gamma\sigma\delta} \nonumber\\
    &\hspace{2cm}+m^2 q_\alpha q_\beta (\tr \gamma^{\mu\alpha\nu}{}_5{}^{\beta\rho\sigma} + \tr \gamma^{\mu\alpha\nu}{}_5{}^{\rho\beta\sigma} + \tr \gamma^{\mu\alpha\nu}{}_5{}^{\rho\sigma\beta}\\
    &\hspace{3.4cm}+ \tr \gamma^{\mu\nu}{}_5{}^{\alpha\rho\beta\sigma} + \tr \gamma^{\mu\nu}{}_5{}^{\alpha\rho\sigma\beta} + \tr \gamma^{\mu\nu}{}_5{}^{\rho\alpha\sigma\beta})\Big] A_\nu D^V_\rho D^V_\sigma \nonumber \,.
\end{align}
There are two other similar contributions with different positions of $\slashed A\gamma_5$.
After evaluation of the traces and the $d+1$ decomposition of both the momenta and the operators, we obtain the structure already described in the massless case
\begin{align}
\begin{split}\label{eq:jVcons}
    \langle\vec j_V\rangle &= 
    f_1\mu_A\vec{B}
    +f_2\vec{A}\times \vec{E}
    +f_3\mu_V\vec{B}_A
    +f_4\vec{V}\times \vec{E}_A \,,\\[3pt]
    \Acal_V[\theta]
    &= \left(f_2+f_3\right) \vec E\cdot \vec B_A +\left(f_1+f_4\right) \vec B\cdot \vec E_A \,,
    \end{split}
\end{align}
except that now the coefficients $f_i$ are non-trivial functions of the dimensionless parameter $(\beta m)$ expressed in terms of the master sum-integrals $S_{a,b,c}$ defined in \eqref{eq:defS}. Their analytical computation in dimensional regularisation is challenging but can be achieved using a Mellin transformation. It introduces an additional integral under which the computation of both the sum and the integral become more manageable. The contour of the remaining integral can be closed and each residue leads to a different order in $(\beta m)$. The result is given as a power series in $\beta m$. More details are available in App.~\ref{app:1}.

Nonetheless, some simplifications can be obtained to all orders in $\beta m$  within the radius of convergence, and then analytically continued outside. In particular we show in Appendix the generic identity~\eqref{eq:idmain} which in particular implies
\begin{align}\label{eq:f0}
    \frac{4}{d}\Big((d-4)S_{0,0,2}+4S_{0,1,3}+4m^2S_{0,0,3}\Big)=0\,,
\end{align}
that we use to verify that the anomaly is independent from $\beta m$ at all orders in the expansion of the $f_i$ functions.
We are also able to reduce all the mass corrections to the vector current to a single function
\begin{align}
    f(\beta m) &= 16m^2 S_{0,0,3} =\frac{m^2}{2\pi^2}\int_0^{\infty} \frac{\dd r\, r^2}{E_r^5}\bigg[3 + 2n_F\Big(-3 + (n_F-1)(3\beta E_r-(2n_F-1)\beta^2E_r^2)\Big)\bigg] \nonumber\\
    &= \frac{7\zeta(3)}{8\pi^4}(\beta m)^2+ \mathcal{O}\left((\beta m)^4\right) \,,\label{eq:f(beta m)}
\end{align}
where $E_r=\sqrt{r^2+m^2}$ and the Fermi distribution is $n_F=\left(1+e^{\beta E_r}\right)^{-1}$.
Eventually we obtain
\begin{align}
    \begin{split}
        f_1(\beta m) 
        & = 0 \,,\\[3pt]
        f_2(\beta m) 
        &= \frac{4}{d}\Big( (3d-8)S_{0,0,2} -4(d-2) S_{0,1,3} +8 m^2S_{0,0,3}\Big)
        =-\frac{1}{2\pi^2}+f(\beta m)\,,\\[3pt]
        f_3(\beta m)&= -S_{0,0,2} +4S_{0,1,3}
        =\frac{1}{2\pi^2}-f(\beta m)\,,\\[3pt]
        f_4(\beta m) 
        & = 0\,,
    \end{split}
\end{align}
where we used~\eqref{eq:f0}, and finally
\begin{align}
\begin{split}\label{eq:jvAvcons}
    \langle\vec j_V\rangle &=
    \left(\frac{1}{2\pi^2}-f(\beta m)\right)\left(\mu_V\vec B_A-\vec{A}\times \vec{E}\right)
    \,,\\     
    \Acal_V[\theta]&= 0 \,.
    \end{split}
\end{align}
The absence of $F_V \tilde F_A$ from the anomaly shows that our regularisation scheme leads to a result satisfying the vector current conservation, even with a mass. This is not surprising as the mass corrections are not expected to influence the UV divergences.
Consequently, the Bloch theorem is still satisfied: the CME remains zero at equilibrium and the AHE and CpME can be combined into a total derivative term.
We however obtain mass corrections to the AHE and the CpME, which to the best of our knowledge, is a new result.

\subsubsection{Axial current}
\begin{figure}
    \centering
    \includegraphics[width=0.8\linewidth]{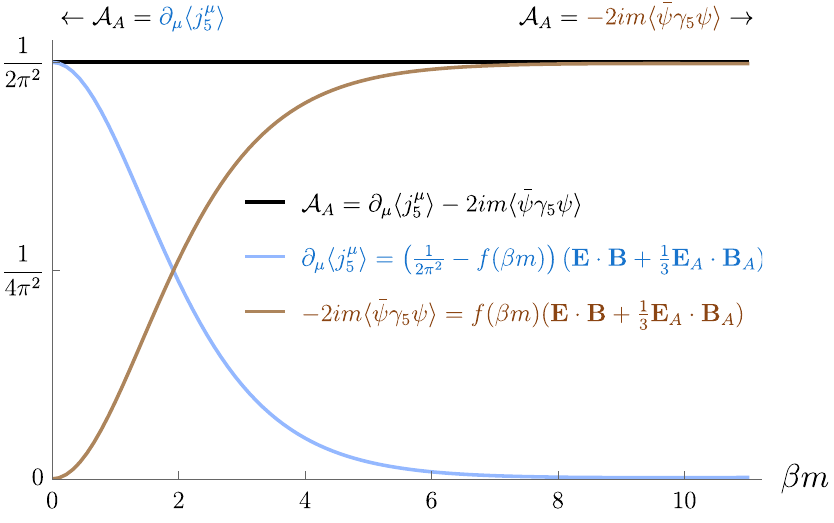}
    \caption{Consistent axial anomaly (black) and its repartition between the divergence of the axial current (blue) and the mass explicit breaking (brown).}
    \label{Fig:plotf}
\end{figure}

As explained in Sec.~\ref{sec:introAxial}, in presence of a mass, the situation changes for the axial symmetry which is now explicitly broken.
The right-hand side of \eqref{eq:CDEjA} with massive propagators still gives access to the current
\begin{align}
\begin{split}
    \langle\vec j_A\rangle &=
    g_1\mu_V\vec{B}
    +g_2\vec{V}\times \vec{E}
    +g_3\mu_A\vec{B}_A
    +g_4\vec{A}\times \vec{E}_A  \,,
\end{split}
\end{align}
where the coefficients are now 
\begin{align}
    \begin{split}\label{eq:defg}
        g_1(\beta m) &=\frac{1}{2\pi^2}-f(\beta m) \,,\\[3pt]
        g_2(\beta m) &=0 \,,\\[3pt]
        g_3(\beta m) &=\frac{1}{3\pi^2}-\frac{2}{3}f(\beta m) \,,\\[3pt]
        g_4(\beta m) &=-\frac{1}{6\pi^2}+\frac{1}{3}f(\beta m)\,.
    \end{split}
\end{align}
Their expressions before reduction is left for App~\ref{app:jAdetail}.
The anomaly still arises from the Jacobian, but due to the explicit breaking from the mass it no longer corresponds to the divergence of the axial current, as explained in Sec.~\ref{sec:introAxial}. The anomaly is thus the difference between the total breaking, and that of the mass given by~\footnote{As emphasised in Sec.~\ref{sec:topol}, we again omit the potential $P$-even contributions in this work.}
\begin{align}
    \begin{split}\label{eq:massterm}
    -2im\vev{\bar\psi\gamma_5\psi}&=-\int_0^{\beta}\!\!\!\!\dd t\int\!\!\dd^3x\,2im\theta\,\frac{1}{\beta}\sum_{n\in\mathds Z}\int\frac{\dd^3 \vec q}{(2\pi)^3}\tr\,\gamma_5\gamma^\mu \sum_{k\in\mathds{N}}\left[\Delta i\sD\right]^k\Delta \\
    &=g_{\rm{mass},1}\vec E\cdot \vec B +g_{\rm{mass},2}\vec B_A\cdot \vec E_A +\mathcal{O}(\beta^2) \\
    &=f(\beta m)\left(\vec E\cdot \vec B +\frac{1}{3}\vec E_A\cdot \vec B_A\right) +\mathcal{O}(\beta^2)
    \,.
    \end{split}
\end{align}
The relevant contributions in \eqref{eq:massterm} arise at $k=4$ where all the sum-integrals are convergent and do not need to be regularised. In particular, there is no ambiguity associated with $\gamma_5$.
Eventually, we obtain the anomaly
\begin{align}
    \begin{split}
    \Acal_A[\theta]
    &=(g_1+g_2+g_{\rm{mass},1})\vec E\cdot \vec B +(g_3+g_4+g_{\rm{mass},2})\vec E_A\cdot \vec B_A \;.
    \end{split}
\end{align}
Finally, injecting the expression for the coefficients $g_i$ and $g_{\mathrm{mass},i}$ we obtain
\begin{align}
\begin{split}
\vev{\vec{j}_A} &= \left(\frac{1}{2\pi^2}-f(\beta m)\right)\left(\mu_V\vec{B}+\frac{2}{3}\mu_A\vec{B}_A-\frac{1}{3}\vec{A}\times \vec{E}_A\right)\;, 
\\
\mathcal{A}_A &=\frac{1}{2\pi^2}\left(\vec E\cdot \vec B +\frac{1}{3}\vec E_A\cdot \vec B_A \right)
\;.
\end{split}
\end{align}
More details are available in App.~\ref{app:jAdetail}.
Once again, the anomaly is found to be independent of $\beta m$.
However, it is interesting here to observe the repartition of the anomaly between the derivative term and the mass term.
On the one hand, in the massless limit, the anomaly is purely the divergence of the axial current $\partial_{\mu}\vev{j_A^{\mu}}$, as computed in the previous section.
On the other hand, in the infinite mass limit, the Matsubara sums become integrals such that we recover the $T=0$ Lorentz invariant master integrals~\cite{Larue:2025yar}. The anomaly is then purely in the pseudo-scalar current $m\vev{\bar\psi\gamma_5\psi}$ appearing in the right-hand side of the Ward identity \eqref{eq:WardA} and the divergence of the axial current vanishes~\cite{Quevillon:2019zrd,Filoche:2022dxl}.
In-between, there is a smooth transition of this repartition given by $f$ in~\eqref{eq:f(beta m)} all the while the anomaly remains constant, see Fig.~\ref{Fig:plotf}. 

A similar analysis can be found in~\cite{Fang:2016uds} in which the distribution of the anomaly between the divergence of the current and the pseudo-scalar term is analysed in both $\beta m$ and $\beta \mu$ directions. However, the anomaly is given and assumed to be constant, the pseudo-scalar term is then inferred from the Ward identity. Only the divergence of the current is computed. On the other hand, in our work, both the divergence of the current and the pseudo-scalar terms are directly evaluated and the anomaly follows from the Ward identity; it is thus proved to be constant. To the best of our knowledge, our analysis of both terms along the $m/T$ flow is new.

Let us compare our results on the transport coefficients with the literature. The original massless computations of the CME and CVE with Kubo formulas in \cite{Kharzeev:2009pj,Landsteiner:2011cp} have been generalised to the massive case in \cite{Lin:2018aon}.
The results we obtained agree with their expression for the mass corrections to the transport coefficients of the CSE. To the best of our knowledge, the mass correction to the other transport effects is a new result.

\section{Scheme-independent anomalies}
\label{sec:EliasAn}

In this Section, we compute the currents and anomalies in a $\gamma_5$-scheme-independent manner. 
In the computation of one-loop divergent triangle Feynman diagrams (associated with the ABJ anomaly), it is well established that, in $4$ dimensions, an ambiguity arises in the loop integral. This ambiguity stems from the arbitrariness in the choice of integration variables~\cite{Weinberg:1996kr}, as surface terms may depend on the momentum routing. These surface terms contribute to the divergence of vector and axial-vector currents, preventing the simultaneous satisfaction of all naive Ward identities. Consequently, at least some of these identities become anomalous. Crucially, the arbitrariness in the integration variables can be parametrised using free parameters, allowing one to select which symmetries are broken at the quantum level and which remain preserved. To ensure physically consistent results, all gauge symmetries must be maintained.

In $4-\epsilon$ dimensions, the ambiguity in loop integrals no longer originates from momentum routing dependencies but instead arises from the Dirac algebra sector. In $ 4-\epsilon$ dimensions, not all properties of Dirac matrices can be preserved, as $\gamma^5$ and the antisymmetric tensor $\epsilon^{\mu\nu\rho\sigma}$ are inherently four-dimensional objects. Regardless of the chosen definition, it is impossible to consistently maintain both the anticommutativity of $\gamma^5$ matrices, i.e., $\{\gamma^\mu,\gamma^5\} = 0$, and the cyclicity of the trace. As noted by 't~Hooft and Veltman\,\cite{tHooft:1972tcz}, the momentum routing ambiguity is replaced by an ambiguity in the placement of $\gamma^5$ within Dirac traces. Their prescription for the Dirac algebra in $4-\varepsilon$ dimensions (see Refs.\,\cite{tHooft:1972tcz,Breitenlohner:1977hr}) introduces free parameters to account for all possible $\gamma^5$ positions in a Dirac matrix string\,\cite{Elias:1982ea}.

The power of this approach is that it yields a master formula depending on free parameters; in particular, we will show that the consistent anomalies can be recovered by enforcing the conservation of the vector symmetry and Bose symmetry of the axial fields, whereas the covariant anomaly follows from enforcing the gauge covariance of the axial and vector currents.
In practice, we will make the replacement \footnote{Note that other positions of $\gamma_5$ could be considered in~\eqref{eq:Elias1} but they are in fact redundant. These four terms are equivalent to the different positions for $A_\mu\gamma_5$ in the CDE expansion.
}
\begin{align}
\begin{split}\label{eq:Elias1}
    \tr \gamma^\mu \gamma^\alpha \gamma^\nu \Red{\gamma_5} \gamma^\beta \gamma^\rho \gamma^\gamma \gamma^\sigma \gamma^\delta 
    \rightarrow
    &a_1 \tr \gamma^\mu \Red{\gamma_5} \gamma^\alpha \gamma^\nu \gamma^\beta \gamma^\rho \gamma^\gamma \gamma^\sigma \gamma^\delta
    +a_2 \tr \gamma^\mu \gamma^\alpha \gamma^\nu \Red{\gamma_5} \gamma^\beta \gamma^\rho \gamma^\gamma \gamma^\sigma \gamma^\delta\\
    +&a_3 \tr \gamma^\mu \gamma^\alpha \gamma^\nu \gamma^\beta \gamma^\rho \Red{\gamma_5} \gamma^\gamma \gamma^\sigma \gamma^\delta
    +a_4 \tr \gamma^\mu \gamma^\alpha \gamma^\nu \gamma^\beta \gamma^\rho \gamma^\gamma \gamma^\sigma \Red{\gamma_5} \gamma^\delta \,,
\end{split}
\end{align}
where the weights $a_i$ are positive real numbers summing to one so as to ensure a consistent $d\to4$ limit.
At this stage, the free coefficients give us the freedom to mimic any possible $\gamma_5$-scheme independently of how the traces are computed. In particular we can compute individual traces using the BMHV-scheme without loss of generality. 
For traces involving three $\gamma_5$'s, the same procedure is applied, keeping only one $\gamma_5$
\begin{align}
\begin{split}\label{eq:Elias2}
    \tr \gamma^\mu \Red{\gamma_5} \gamma^\alpha \gamma^\nu \Red{\gamma_5} \gamma^\beta \gamma^\rho \Red{\gamma_5} \gamma^\gamma \gamma^\sigma \gamma^\delta 
    \rightarrow
    &a_1 \tr \gamma^\mu \Red{\gamma_5} \gamma^\alpha \gamma^\nu \gamma^\beta \gamma^\rho \gamma^\gamma \gamma^\sigma \gamma^\delta
    +a_2 \tr \gamma^\mu \gamma^\alpha \gamma^\nu \Red{\gamma_5} \gamma^\beta \gamma^\rho \gamma^\gamma \gamma^\sigma \gamma^\delta\\
    +&a_3 \tr \gamma^\mu \gamma^\alpha \gamma^\nu \gamma^\beta \gamma^\rho \Red{\gamma_5} \gamma^\gamma \gamma^\sigma \gamma^\delta
    +a_4 \tr \gamma^\mu \gamma^\alpha \gamma^\nu \gamma^\beta \gamma^\rho \gamma^\gamma \gamma^\sigma \Red{\gamma_5} \gamma^\delta \,.
\end{split}
\end{align}
The number of coefficients introduced will be larger than the number of constrains; however, most of them will be found to contribute in the same way or not at all, so that effectively the number of coefficients is greatly reduced.

\subsection{Master formulae}
Let us focus on the changes in the computation for the vector current, i.e. how the free parameters affect the result. We will consider again the massive case. Importantly, let us emphasise that the mass corrections arise from UV and IR finite integrals. As a result, they are non-ambiguous hence independent from the free parameters. This means that we only expect the free parameters to appear at order $(\beta m)^0$ in the results.
Let us return to the expressions Eqs.~\eqref{eq:jVk3},~\eqref{eq:jVex} and introduce different sets of four free parameters for each trace of the form \eqref{eq:Elias1}.
Three such sets are needed, denoted $a_i,b_i,c_i$ and satisfying $x_i\in[0,1],\,\sum_i x_i=1,\,x\in\{a,b,c\}$.
We thus start with nine parameters, although after computation the result depends on only two linear combinations of them $\kappa_1\in[0,1],\,\kappa_2\in[0,2]$ such that
\begin{align}
    \begin{split}\label{eq:MasterV}
    \langle\vec j_{V}\rangle &=
    \left(\frac{\kappa_2-\kappa_1+1}{4\pi^2}\right)\mu_A\vec{B}
    +\left(\frac{\kappa_2-\kappa_1-1}{4\pi^2}+f(\beta m)\right)\vec{A}\times \vec{E}\\
    &\;\,+\left(\frac{\kappa_1+1}{4\pi^2}-f(\beta m)\right)\mu_V\vec{B}_A
    +\frac{\kappa_1-1}{4\pi^2}\vec{V}\times \vec{E}_A \,,\\
    \Acal_{V}[\theta]
    &= \frac{\kappa_2}{4\pi^2}(\vec E\cdot \vec B_A +\vec B\cdot \vec E_A) \,,
\end{split}
\end{align}
where the mass corrections involve $f$ given in~\eqref{eq:f(beta m)}. 

Turning to the axial current, the contributions from the vector field are the same as in the consistent anomaly~\eqref{eq:defg}. The only modifications occurs on the terms that involve three $\gamma_5$'s. After further manipulations, the axial current and anomaly turn out to depend on a single linear combination of free parameters that we relabel as $\kappa\in[0,1]$, such that
\begin{align}
\begin{split}\label{eq:MasterA}
    \langle\vec j_A\rangle &=
    \left(\frac{1}{2\pi^2}-f(\beta m)\right)\mu_V\vec{B}+\left(\frac{\kappa+1}{4\pi^2}-\frac{2}{3}f(\beta m)\right)\mu_A\vec{B}_A\\
    &+\left(\frac{\kappa-1}{4\pi^2}+\frac{1}{3}f(\beta m)\right)\vec{A}\times \vec{E}_A \,.\\
    \Acal_{A}[\theta]
    &= \frac{1}{2\pi^2} (\vec E\cdot \vec B + \kappa\, \vec E_A\cdot \vec B_A)\,.
\end{split}
\end{align}
We recall that the axial anomaly is obtained by subtracting the explicit breaking from the mass $2im\vev{\bar\psi\gamma_5\psi}$ which is UV finite hence independent from any free parameters. Note that here our use of the Elias trick is automatically compatible with the ABJ anomaly~\cite{Adler:1969gk,Bell:1969ts}. Indeed, if one takes $\vec A=0$ and $\mu_A=0$, then the free parameters do not contribute anymore.

Let us clarify a point. The anomalous currents have the same cancellation condition than the anomalies: $\sum_L q_L^3-\sum_R q_R^3=0$ where $q_{R/L}$ are the $U(1)_{L/R}$ charges of left-/right-handed fermions. However, the vanishing of an anomaly does not always imply the vanishing of the corresponding anomalous current. For example, although the consistent vector anomaly vanishes~\eqref{eq:AVm0cons}, the AHE and CpME are still generated~\eqref{eq:jVm0cons} due to the non-vanishing of the axial anomaly coefficient. Likewise, the anomaly operators vanish in the absence of electric fields at global equilibrium, nevertheless the CSE, CpSE, CME and CpME are still generated.

Eqs.~\eqref{eq:MasterV} and~\eqref{eq:MasterA} are our master formulae from which we can recover the consistent anomaly from the previous Section, as well as the covariant anomaly as we will show below.

\subsection{Consistent anomaly}
\label{sec:consistentkappa}

The consistent vector current and anomaly are easily recovered by requiring the conservation of the vector symmetry. Firstly, it immediately sets $\kappa_2=0$ in Eq.~\eqref{eq:MasterV} such that $\mathcal{A}_V=0$. Secondly, the vector current should depend on $\textbf{V}$ only through $F_V$ which implies that $\kappa_1=1$ in Eq.~\eqref{eq:MasterV}. We thus recover the  consistent vector current and anomaly Eq.~\eqref{eq:jvAvcons}.

In the abelian case both the covariant and the consistent anomalies are Wess-Zumino consistent (see e.g. footnote~\ref{footnote1}). Following~\cite{Filoche:2022dxl}, one could compute the anomalies in the non-abelian case in a $\gamma_5$-scheme independent manner, fix the free parameters by imposing the Wess-Zumino consistency conditions, and then take the abelian limit. The same result can alternatively be recovered without resorting to the non-abelian case by imposing Bose symmetry of the axial fields which can be easily seen at the level of the anomaly $\mathcal{A}_A$. The first term $\vec E\cdot\vec B$ corresponds to a $\partial j_A VV$-diagram, and the vector current conservation implies that all the anomaly is in the only axial leg $\partial j_A$. On the other hand, the second term $\vec E_A\cdot\vec B_A$ corresponds to a $\partial j_AAA$-diagram. Bose symmetry implies that  the anomaly is evenly distributed among its three legs, such that the axial current violation in the $\partial j_A A A$ diagram is only 1/3 that of the $\partial j_A V V$~\cite{Bertlmann:1996xk}, i.e. $\kappa=1/3$.

\subsection{Covariant anomaly}
\label{sec:covariantkappa}

In the massless limit, the currents are made covariant simply by requiring that they only depend on $\textbf{V}$ and $\textbf{A}$ through the fields strengths. It is immediate from Eqs.~\eqref{eq:MasterV} and ~\eqref{eq:MasterA} that this implies: $\kappa_1=1$, $\kappa_2=2$ and $\kappa=1$. Note that in the massive case, it is not possible to make the axial current covariant since the axial symmetry is classically broken.

Eventually, we obtain the covariant vector current and anomaly
\begin{align}
    \begin{split}
    \langle\vec j_{V,\rm cov}\rangle &=
    \frac{1}{2\pi^2}\mu_A\vec{B}+f(\beta m)\vec{A}\times \vec{E}+\left(\frac{1}{2\pi^2}-f(\beta m)\right)\mu_V\vec{B}_A \,,\\
    \Acal_{V,\rm cov}[\theta]
    &= \frac{1}{4\pi^2}F_V\tilde F_A\,,
    \end{split}
\end{align}
and the covariant axial current and anomaly
\begin{align}
\begin{split}
    \langle\vec j_{A,\rm cov}\rangle &=
    \frac{1}{2\pi^2}\mu_V\vec{B}
    +\left(\frac{1}{2\pi^2}-\frac{2}{3}f(\beta m)\right)\mu_A\vec{B}_A +\frac{1}{3}f(\beta m)\vec{A}\times \vec{E}_A \,,\\[3pt]
    \Acal_{A,\rm cov}[\theta]
    &= \frac{1}{2\pi^2} (\vec E\cdot \vec B + \vec E_A\cdot \vec B_A) = \frac{1}{8\pi^2} (F_V\tilde F_V+F_A\tilde F_A) \,.
    \end{split}
\end{align}
We note that the differences between the covariant and consistent currents are precisely the abelian Bardeen-Zumino polynomials~\cite{Bardeen:1984pm, Bertlmann:1996xk}: $j^\mu_{V,BZ} =\frac{1}{2\pi^2} \epsilon^{\mu\nu\rho\sig}A_\nu\partial_\rho V_\sigma$ and $j^\mu_{A,BZ}=\frac{1}{6\pi^2}\epsilon^{\mu\nu\rho\sig}A_\nu\partial_\rho A_\sigma$.

It is interesting to see that we now obtain the CME $\mu_A\vec  B$. This is not surprising since the vector gauge symmetry is now violated, and therefore the Bloch theorem does not apply (see also the discussion in \cite{Landsteiner:2016led, Gorbar:2016ygi}). 

The expressions of the covariant and consistent currents and anomalies are summarised in Tables~\ref{tab:vector} and~\ref{tab:axial}.

\begin{table}[h!]
\centering

\begin{subtable}
\centering
\renewcommand{\arraystretch}{1.5}
\setlength{\tabcolsep}{14pt}

\begin{tabular}{|c|cccc|}\hline
Chern-Simons & \multicolumn{2}{c|}{$\epsilon^{\mu\nu\rho\sigma}A_\nu\partial_\rho V_\sigma$} &\multicolumn{2}{c|}{$\epsilon^{\mu\nu\rho\sigma}V_\nu\partial_\rho A_\sigma$} \\\hline
3+1 separation & $\mu_A \vec B$ & $\vec A \times \vec E$ & $\mu_V \vec B_A$ & $\vec V \times \vec E_A$ \\
Effect & CME \cite{Fukushima:2008xe} & AHE \cite{Zyuzin:2012tv} & CpME \cite{Grushin:2016fgr} & /  \\\hline
$\langle j_{V,\mathrm{cons}}\rangle$ & 0 & $-\frac{1}{2\pi^2}+f(\beta m)$ & $\frac{1}{2\pi^2}-f(\beta m)$ & 0 \\ [2ex]
$\Acal_{V,\mathrm{cons}}$&\multicolumn{4}{c|}{0} \\[5pt] \hline
$\langle j_{V,\mathrm{cov}}\rangle$ & $\frac{1}{2\pi^2}$ & $f(\beta m)$ & $\frac{1}{2\pi^2}-f(\beta m)$ & 0 \\ [2ex]
$\mathcal{A}_{V,\mathrm{cov}}$ & \multicolumn{4}{c|}{$\displaystyle \frac{1}{2\pi^2} \left( E \cdot B_A + E_A \cdot B \right)=\frac{1}{4\pi^2} F_V\tilde F_A$}\\[5pt] \hline
\end{tabular}

\caption{Summary table for the consistent and covariant vector current and anomaly (their divergence) at finite mass and temperature. Mass and temperature corrections are captured in the function $f(x)  = \frac{7\zeta(3)}{8\pi^4}x^2+ \mathcal{O}\left(x^4\right)\underset{x\rightarrow\infty}{\rightarrow} \frac{1}{2\pi^2}$ defined in \eqref{eq:f(beta m)}.}
\label{tab:vector}

\end{subtable}

\begin{subtable}
\centering
\renewcommand{\arraystretch}{1.5}
\setlength{\tabcolsep}{11.3pt}

\begin{tabular}{|c|cc|cc|}\hline
Chern-Simons & \multicolumn{2}{c|}{$\epsilon^{\mu\nu\rho\sigma}V_\nu\partial_\rho V_\sigma$} & \multicolumn{2}{c|}{$\epsilon^{\mu\nu\rho\sigma}A_\nu\partial_\rho A_\sigma$} \\\hline
3+1 separation& $\mu_V \vec B$ & $\vec V \times \vec E$ & $\mu_A \vec B_A$ & $\vec A \times \vec E_A$ \\
Effect& CSE \cite{Metlitski:2005pr} & / & CpSE \cite{Huang:2017rpa} & AAHE \cite{Huang:2017rpa} \\\hline
$\langle j_{A,\mathrm{cons}}\rangle$ & $\frac{1}{2\pi^2}-f(\beta m)$ & 0 & $\frac{1}{3\pi^2}-\frac{2}{3}f(\beta m)$ & $-\frac{1}{6\pi^2}+\frac{1}{3}f(\beta m)$ \\ [2ex]
$\Acal_{A,\mathrm{cons}}$ & \multicolumn{2}{c|}{$\frac{1}{2\pi^2}E \cdot B = \frac{1}{8\pi^2}F_V \tilde F_V$} & \multicolumn{2}{c|}{$\frac{1}{6\pi^2}E_A \cdot B_A = \frac{1}{3}\frac{1}{8\pi^2}F_A\tilde F_A$} \\[5pt] \hline
$\langle j_{A,\mathrm{cov}}\rangle$ & $\frac{1}{2\pi^2}-f(\beta m)$ & 0 & $\frac{1}{2\pi^2}-\frac{2}{3}f(\beta m)$ & $\frac{1}{3}f(\beta m)$ \\ [2ex]
$\Acal_{A,\mathrm{cov}}$ & \multicolumn{2}{c|}{$\frac{1}{2\pi^2}E \cdot B =\frac{1}{8\pi^2} F_V\tilde F_V$} & \multicolumn{2}{c|}{$\frac{1}{2\pi^2}E_A \cdot B_A =\frac{1}{8\pi^2} F_A\tilde F_A$} \\[5pt] \hline
\end{tabular}

\caption{Summary table for the consistent and covariant axial current and anomaly at finite mass and temperature. Contrary to the vector case, the anomaly is obtained from the difference of the divergence of the current and the explicit symmetry breaking term due to the mass~\eqref{eq:WardA}. Mass and temperature corrections are given by $f(x)  = \frac{7\zeta(3)}{8\pi^4}x^2+ \mathcal{O}\left(x^4\right)\underset{x\rightarrow\infty}{\rightarrow} \frac{1}{2\pi^2}$ defined in \eqref{eq:f(beta m)}.}
\label{tab:axial}

\end{subtable}
\end{table}

\subsection{Chern-Simons currents and $m/T$-flow}
\label{sec:m/Tflow}
Although it is well-known that the Pontryagin densities that constitute the anomalies~\eqref{eq:MasterV} and~\eqref{eq:MasterA} are the divergence of the corresponding Chern-Simons currents, it is not obvious that the physical currents themselves correspond to these Chern-Simons.

Take for example the simple case of the abelian anomaly without axial background, mass nor temperature. The vector symmetry is preserved $\partial\vev{j_V}=0$ whereas the axial global symmetry is broken $\partial\vev{j_A}=\frac{1}{8\pi^2} F\tilde F$~\cite{Adler:1969gk}. Since the anomaly is the divergence of the Chern-Simons current $F\tilde F=\partial_\mu \left(\epsilon^{\mu\nu\rho\sigma}V_\nu F_{V,\rho\sigma}\right)$, it is tempting to identify the physical current $\vev{j_A}$ with it. The Chern-Simons current is however not locally gauge covariant, this would therefore violate the vector symmetry of the theory and contradict $\partial\vev{j_V}=0$.~\footnote{Using Bianchi identity, the Chern-Simons current is gauge invariant up to total derivative $\partial_\mu(\epsilon^{\mu\nu\rho\sigma}\theta_V F_{\rho\sigma})$, but $\partial\vev{j_V}=0$ implies local vector gauge invariance.}
As we will see, this apparent contradiction is resolved when the theory is properly IR regulated.
In fact, an important result of this paper is that the physical currents only correspond to the Chern-Simons in the $m/T\to\infty$ limit. To better understand this result, let us investigate how the currents flow between the $m/T\gg1$ and $m/T\ll1$ limits.

In this work, we obtained the vector and axial currents and anomalies at finite mass and temperature. They are expressed in terms of a function $f(m/T)$ defined in Eq.~\eqref{eq:f(beta m)} and plotted in Fig.~\eqref{Fig:plotf}. Although we only have an integral expression for $f(m/T)$ for all $m/T$, an analytic form can be achieved in the $m/T\ll1$ limit and is expressed as a power series in $m/T$. Additionally, the $m/T\gg 1$ limit is easily obtained at leading order in $m/T$ as $f(x)\underset{x\to\infty}{\longrightarrow}\frac{1}{2\pi^2}$.
Interestingly, although the Lorentz symmetry is broken due to the preferred time-direction inherent to having a non-zero temperature, it is expected to be restored in the $m/T\to\infty$ since the temperature dependence drops, and therefore the vacuum result should be recovered~\cite{Larue:2025yar}. Let us stress that although the anomalies are independent from $m/T$ owing to their topological nature, the physical currents from Eqs.~\eqref{eq:MasterV} and~\eqref{eq:MasterA} are not, and in particular they are not Lorentz-covariant at non-zero temperature.

Regardless, in the literature~\cite{Zyuzin:2012tv,Chen:2013mea, Hosur:2013kxa,Huang:2017rpa} the physical currents are often identified with the Chern-Simons currents
\begin{equation}
j^\mu_{\mathrm{CS},XY}\equiv\epsilon^{\mu\nu\rho\sigma}X_\nu\partial_\rho Y_\sigma\;,\quad\quad X,Y\in\{V,A\}\;,
\end{equation}
without mass or temperature considerations. This is done by assuming the covariant form of the anomalies and having the currents divergences reproduce them.~\footnote{A more careful approach can be found in one of the early derivations of the CME in \cite{Fukushima:2008xe}. The covariant vector current is obtained to be $j_{\mathrm{CS},AV}$, from which the CME is obtained by specifying $A_0=\mu_A$. In this context, a mass is kept but the result is argued to be independent of it, contrary to the axial current which has an explicit breaking. This result agrees with ours, i.e. the $m/T\rightarrow\infty$ limit of our covariant current.}
However, these currents are Lorentz-covariant. Crucially, this implies that the physical currents in~\eqref{eq:MasterV} and~\eqref{eq:MasterA} cannot be reconstructed from the Chern-Simons at non-zero temperature.~\footnote{Additionally,  in the absence of temperature, the identification of time-component of the gauge fields with the chemical potentials is dubious since it can be gauged-away~\cite{Landsteiner:2012kd}.}

To illustrate this, let us consider for example $\vec j_{\mathrm{CS},AA}$. At finite temperature, with $A_0=\mu_A$ and $\partial_0 A_\mu=0$, it splits into two operators $\vec j_{\mathrm{CS},AA}=\mu_A\vec B_A+\vec A\times \vec E_A$. However, one can see from Eq.~\eqref{eq:MasterA} that at non-zero temperature these two operators come with different coefficients. It is only in the $m/T\to\infty$ limit that both coefficients flow into the same such that the Lorentz-covariant Chern-Simons current is recovered.   
The same applies for the other Chern-Simons currents using $V_0=\mu_V$, $A_0=\mu_A$ and vanishing time-derivatives
\begin{alignat}{3}
&\#_1\, \mu_V \vec B &\;+\;& \#_2\,\vec V\times \vec E&\;\xrightarrow[\beta m\to\infty]{}\;& j^i_{\mathrm{CS},VV}\;,\nonumber\\
&\#_3\,\mu_A \vec B_A &\;+\;& \#_4\, \vec A\times \vec E_A& \;\xrightarrow[\beta m\to \infty]{}\;& j^i_{\mathrm{CS},AA}\;,\label{eq:CScurrents}\\
&\#_5\, \mu_V \vec B_A &\;+\;& \#_6\, \vec V\times \vec E_A &\;\xrightarrow[\beta m\to\infty]{}\;& j^i_{\mathrm{CS},VA}\;,\nonumber\\
&\#_7\,\mu_A \vec B &\;+\;& \#_8\,\vec A\times \vec E& \;\xrightarrow[\beta m\to\infty]{}\;& j_{\mathrm{CS},AV}\;,\nonumber
\end{alignat}
where $\#_i$ are coefficients that depend on $\beta m$ and can be read from~\eqref{eq:MasterV} and~\eqref{eq:MasterA}. The physical currents thus become Lorentz-covariant and are expressed in terms of the Chern-Simons
\begin{align}
\begin{split}
&\vev{ \vec{j}_V}\underset{\beta m\to\infty}{=}\left(\frac{\kappa_1-\kappa_2-1}{4\pi^2}\right)\vec j_{\mathrm{CS},AV}+\left(\frac{1-\kappa_1}{4\pi^2}\right)\vec j_{\mathrm{CS},VA}\;,\\
&\vev{ \vec{j}_A}\underset{\beta m\to\infty}{=}\left(\frac{1/3-\kappa}{4\pi^2}\right)\vec j_{\mathrm{CS},AA}\;.\label{eq:jlimCS}    
\end{split}
\end{align}
Let us return to the example of the abelian anomaly in the absence of axial background discussed at the beginning of this subsection. One can see from~\eqref{eq:jlimCS} how the contradiction is resolved: since the only Chern-Simons available $j_{\mathrm{CS},VV}$ violates the vector symmetry, it cannot enter the physical current $\vev{j_A}$, and in fact, the physical vector and axial currents are both vanishing. This does not imply the vanishing of the axial anomaly since the breaking from the mass has to be subtracted appropriately (see Fig.~\ref{Fig:plotf})~\cite{Larue:2023uyv}. 
Likewise, in the presence of axial background, the vector symmetry is again preserved for the consistent currents ($\kappa_1=1$, $\kappa_2=0$ and $\kappa=1/3$) in which case both physical currents vanish as well in this limit.

The take-away of this discussion is that although anomalies are independent from the IR regularisation scale, the anomalous currents are not and require a careful IR regularisation  (either by a mass or temperature). A common mistake in the literature is to identify them with the Chern-Simons currents, which is not correct in general.

\section{Conclusion}\label{sec:conclusion}
In this article, we have integrated out a Dirac fermion with vector and axial background at finite temperature and mass at local equilibrium in the adiabatic regime. We have obtained the $P$-odd contributions to transport in the vector and axial currents up to operators of mass dimension three, as well as the associated anomalies.
The computation is fully self-contained and no zero-temperature result is pre-supposed.

The effective theory approach allows us to obtain all at once well-known and sought for effects (CME, CSE, AHE), as well as less studied ones (CpME, CpSE, AAHE) and their mass corrections. The latter effects are less common mainly due to the necessity of an axial background $\vec A$, whose introduction requires more justification and particular care to the regularisation. To the best of our knowledge, they have only been postulated from an identification of the transport currents with the Chern-Simons currents (which is wrong in general as discussed below), and only directly evaluated for the first time in this paper.
The transport coefficients are collected in Tables~\ref{tab:vector} and \ref{tab:axial}. It is clearly established how each transport effect directly arises from a term in the anomaly, or how different contributions conspire to cancel and not enter the anomaly at all. In particular, one may appreciate that anomalous transport currents can still be generated even when the anomaly vanishes. We believe our work helps bridge a gap in the literature by clarifying the precise relationship between quantum anomalies and anomalous transport.

To achieve these results, we first carry out  a full treatment of the UV-regularisation of the consistent currents and anomalies within dimensional regularisation. We then carry out a computation while keeping control over which symmetries are enforced, leading to our master formulae Eqs.~\eqref{eq:MasterV} and~\eqref{eq:MasterA}. The consistent and covariant forms of the currents and anomalies are particular cases of these master formulae. 

The IR-regularisation is also a central point of the computation of the chiral anomaly. In our computation, the IR-finiteness is ensured by the temperature or the mass, with well-defined $m/T\to\infty$ and $m/T\to 0$ limits. Importantly, the introduction of the IR regulators $m$ and $T$ explicitly breaks the axial and  Lorentz symmetries respectively. The dimensionless parameter $m/T$ monitors the breaking of either symmetry. Our computation allows to smoothly analyse the flow between each limit, and see the restoration of the axial symmetry for $m/T\to0$. Crucially, we also show that the transport vector and axial currents are expressed in terms of Chern-Simons currents only in the $m/T\to\infty$ limit where the Lorentz symmetry is restored. In general, the transport currents cannot be expressed in terms of Chern-Simons at finite temperature, as is often assumed in the literature.

Contributions related to non-topological anomalies can also induce anomalous hydrodynamical transport. A systematic study of non-topological contributions (both $P$-even and $P$-odd) will be presented in \cite{Larue:2026}. 

\section*{Acknowledgments}
 We thank Pham Ngoc Hoa Vuong for helpful discussions, and Maxim Chernodub, Karl Landsteiner and Adolfo Grushin for valuable comments on the draft. The work of A.M, J.Q and D.S. is supported by the EFFORT project, funded through the IRGA program of Université Grenoble Alpes (UGA), and by the Tremplin project from CNRS Physique. The work of R.L is supported by the Science and Technology Commission of Shanghai Municipality (grant No. 24ZR1450600).

\appendix

\section{Computational details}\label{app:1}
In this Appendix, we detail the computation of the master sum-integrals, axial and vector currents and anomalies.

\subsection{Reducion to master sum-integrals}
The terms that arise in our computations are of the form
\begin{align}
    \frac{1}{\beta}\sum_n \int\frac{\dd^d\vec q}{(2\pi)^d} \frac{q_{\alpha_1} \dots q_{\alpha_r}}{(\omega_n^2+\vec q^2+m^2)^c}t^{(r)}_{\alpha_1\dots\alpha_r}{}^{\mu\nu\rho\sigma}\,,
\end{align}
where $t^{(r)}$ denotes a Dirac trace with one $\gamma_5$ and $4+r$ Dirac matrices, which is discussed further below.
Because of the discrete sum,  it is not possible to do the usual replacement $q_{\alpha_1}\dots q_{\alpha_{2n}} \rightarrow q^{2n}\left( \prod_{i=0}^{n-1} (d+2i) \right)^{-1}g_{\alpha_1\dots \alpha_{2n}}$ where $g_{\alpha_1\dots \alpha_{2n}}$ is the fully symmetrised metric without normalisation, which requires Lorentz invariance. Instead, we decompose $q$ and use do this replacement only on the spatial piece
\begin{align}
    \begin{split}
    q_\alpha q_\beta &=\delta^0_\alpha\delta^0_\beta \omega_n^2 + \delta^i_\alpha\delta^j_\beta q_i q_j\\
    &\rightarrow \delta^0_\alpha\delta^0_\beta \omega_n^2 + \delta^i_\alpha\delta^j_\beta \vec q^2 \frac{1}{d}\delta_{ij}\\
    q_\alpha q_\beta q_\gamma q_\delta
    &= \delta^0_\alpha\delta^0_\beta \delta^0_\gamma\delta^0_\delta \omega_n^4 + (\delta^0_\alpha\delta^0_\beta\delta^i_\gamma\delta^j_\delta + \dots )  \omega_n^2 q_i q_j + \delta^i_\alpha\delta^j_\beta \delta^k_\gamma\delta^l_\delta q_i q_j q_k q_l\\
    &\rightarrow \delta^0_\alpha\delta^0_\beta \delta^0_\gamma\delta^0_\delta \omega_n^4 + (\delta^0_\alpha\delta^0_\beta\delta^i_\gamma\delta^j_\delta + \dots )  \omega_n^2 \vec q^2 \frac{1}{d}\delta_{ij}\\&\;+ \delta^i_\alpha\delta^j_\beta \delta^k_\gamma\delta^l_\delta \vec q^4 \frac{1}{d(d+2)}(\delta_{ij}\delta_{kl}+\delta_{ik}\delta_{jl}+\delta_{il}\delta_{jk})
    \end{split}
\end{align}
This procedure yields scalar integrals which are easier to compute.

\subsection{Evaluation of the master sum-integrals}
Our goal is to reduce the sum-integrals to the master sum-integral
\begin{align}
    S_{a,b,c} = \frac{1}{\beta}\sum_{n\in\mathds Z}\int\frac{\dd^d\vec q}{(2\pi)^d}\frac{\vec q^{2a} \omega_n^{2b}}{(\omega_n^2+\vec q^2 + m^2)^c} \,.
\end{align}
Throughout the computation we set $d=3-\epsilon$. We are particularly interested in the values of $a$, $b$ and $c$ for which these sum-integrals are divergent in the $\eps\to0$ limit. Although the poles $1/\eps$ turn out to cancel in the anomalous currents and anomalies, they yield finite contributions of the form $\eps/\eps$ which are crucial to account for.

We first start by lowering $a$ to zero iteratively using
\begin{align}
    S_{a,b,c} = S_{a-1,b,c-1}-S_{a-1,b+1,c}-m^2 S_{a-1,b,c} \;,\quad\text{for  }\, a>0\;.
\end{align}
The evaluation of either the sum or the integral is direct, but doing both is more challenging.
Yet, doing so is essential to obtain an analytic expression of the divergence. Let us adapt the derivation based on the introduction of the Mellin transform \cite{Bedingham:2000ct, Larue:2025yar} to the present case (massive fermion).
Using the symmetry of the Matsubara frequencies, we start by restricting the sum to positive integers in exchange of a factor of 2. We also consider the integral to be a function of the Matsubara frequencies
\begin{equation}
    S_{0,b,c}=\frac{1}{\beta}\sum_n I_{b,c}(\omega_n) \,,\quad I_{b,c}(\omega_n) = \int \frac{d^d\vec q}{(2\pi)^d} \frac{\omega_n^{2b}}{(\vec q^2 + \omega_n^2 + m^2)^c}\,,
\end{equation}
 such that we can introduce its inverse Mellin transform
\begin{equation}
    I_{b,c}(\omega_n) = \frac{1}{2\pi i} \int_{r-i\infty}^{r+i\infty} ds \ \omega_n^{-s} \mathcal{M}[I_{b,c}, s]\,,
\end{equation}
with $r\to1^+$.
Substituting this back into our expression and commuting the sum and the integral, we obtain
\begin{equation}
    S_{0,b,c} = \frac{1}{2\pi i \beta} \int_{\mathcal{C}} ds \left( 2 \sum_{n\in \mathds{N}} \omega_n^{-s} \right) \mathcal{M}[I_{b,c}, s] \,.
\end{equation}
We now evaluate the two components of the integrand separately. On one hand, evaluating the sum where $\omega_n = (2n+1)\pi T$ gives
\begin{equation}
    2 \sum_{n\in\mathds{N}} \omega_n^{-s} = 2 \left( \frac{\pi}{\beta} \right)^{-s} (1 - 2^{-s}) \zeta(s) \,.
\end{equation}
On the other hand, the Mellin transform of $I_{b,c}(y)$ combines with the momentum integral to increase its dimension
\begin{align}
    \mathcal{M}[f, s] &= \int_0^\infty dy \ y^{s-1} \int \frac{d^d\vec q}{(2\pi)^d} \frac{y^{2b}}{(\vec q^2 + y^2 + m^2)^c} \nonumber \\
    &= \frac{\Gamma(\frac{s}{2}+b)}{2\pi^{\frac{s}{2}+b}}(2\pi)^{s+2b}\int \frac{d^{d+s+2b}\vec q}{(2\pi)^{d+s+2b}} \frac{1}{(\vec q^2 + m^2)^c}\\
    &= \frac{\Gamma\left(\frac{s}{2}+b\right)}{2(4\pi)^{d/2}} m^{-(2(c-b)-d-s)} \frac{\Gamma\left(c-b-\frac{d}{2}-\frac{s}{2}\right)}{\Gamma(c)} \nonumber
\end{align}
Substituting these results back into the contour integral yields
\begin{align}
\begin{split}
    S_{0,b,b}=&m^{-(2(c-b)-d-s)}\frac{(4\pi)^{-d/2}}{\pi \Gamma(c)}\\
    &\times\frac{1}{2\pi i} \int_{\mathcal{C}} ds \ (1-2^{-s}) \zeta(s) \Gamma\left(\frac{s}{2}+b\right) \Gamma\left(c-b-\frac{d}{2}-\frac{s}{2}\right) \left( \frac{m\beta}{\pi} \right)^{s-1}\;.
\end{split}
\end{align}
We evaluate this integral by closing the contour in the right half-plane and applying Cauchy's residue theorem. The poles are located where the argument of $\Gamma\left(c-b-\frac{d}{2}-\frac{s}{2}\right)$ is a non-positive integer $-k$ for $k \in \mathbb{N}$:
\begin{equation}
    s_k = 2(c-b) - d + 2k \,.
\end{equation}
Near these poles, we use the expansion $\Gamma\left(-k - \frac{x}{2}\right) \sim -2 \frac{(-1)^k}{k!} \frac{1}{x}$.
Summing the residues, we arrive at our final analytic result
\begin{align}
\begin{split}
    S_{0,b,c}&=\frac{2^{1-d}}{\pi\Gamma (c)}m^{1-s_0}\sum_{k\in\mathds N} \frac{(-1)^k}{k!} \left(1-2^{-s_k}\right) \pi ^{-\frac{d}{2}-s_k} \Gamma \left(c-\frac{d}{2}+k\right) \zeta (s_k) (\beta  m)^{s_k-1}\,.
    \end{split}
\end{align}
The UV-divergences we were looking for appear in the limit $\zeta(1+\eps)\rightarrow\frac{1}{\eps}+\gamma_E$ when $d=3-\eps\rightarrow3$.
Note that when evaluating $\zeta$ with even lower arguments, we can use its analytic continuation, in particular $\zeta(-1)=-1/12$. This does not concern the sum-integrals used in this article, but it is encountered in the computation of the temperature contribution to the Chiral Vortical Effect~\cite{Larue:2025yar}.
The evaluation of the sum-integrals for the first values of $b,\, c$ can be found in Table \ref{tab:sum-integrals}.
\begin{table}[h!]
\centering
\begin{tabular}{|c|c|c|c|c|}
\hline
$c \backslash b$
 & 0 & 1 & 2 \\ \hline
1 & $\frac{1}{\beta ^2}\left(-\frac{1}{24}-\frac{(\beta m)^2}{8 \pi^2 \epsilon} + ..\right)$ & $\frac{1}{\beta ^4}\left(\frac{7 \pi^2}{240} +.. + \frac{(\beta m)^4}{32 \pi^2 \epsilon} + ..\right)$ & $\frac{1}{\beta ^6}\left(-\frac{31 \pi^4}{504} + .. - \frac{(\beta m)^6}{64 \pi^2 \epsilon} + ..\right)$ \\ \hline
2 & $\frac{1}{8 \pi^2 \epsilon} +\frac{r}{16 \pi^2 } + ..$ & $\frac{1}{\beta ^2}\left(\frac{1}{48} - \frac{(\beta m)^2}{16 \pi^2 \epsilon} + ..\right)$ & $\frac{1}{\beta ^4}\left(-\frac{7 \pi^2}{480} +.. + \frac{3(\beta m)^4}{64 \pi^2 \epsilon} + ..\right)$ \\ \hline
3 & $\beta^2\left(\frac{7\zeta(3)}{128 \pi^4} + ..\right)$ & $\frac{1}{32 \pi^2 \epsilon} +\frac{r+2}{64 \pi^2} + ..$ & $\frac{1}{\beta ^2}\left(\frac{1}{192 } - \frac{3(\beta m)^2}{64 \pi^2 \epsilon} + ..\right)$ \\ \hline
4 & $\beta^4\left(\frac{31 \zeta(5)}{1024 \pi^6} + ..\right)$ & $\beta^2\left(\frac{7 \zeta(3)}{256 \pi^4} + ..\right)$ & $\frac{1}{64 \pi^2 \epsilon} +\frac{3r + 8}{384 \pi^2 } + ..$ \\ \hline
5 & $\beta^6\left(\frac{635\zeta(7)}{32768 \pi^8} + ..\right)$ & $\beta^4\left(\frac{155\zeta(5)}{8192 \pi^6} + ..\right)$ & $\beta^2\left(\frac{35 \zeta(3)}{2048 \pi^4} + ..\right)$ \\ \hline
\end{tabular}
\caption{$S_{0,b,c}$ to leading orders  in $\beta m$ with $d\rightarrow3-\eps$, while still displaying the pole $1/\eps$. $r=\gamma_E+\log4/\pi +\log\beta^2\mu_{\mathrm{ren}}^2$ with $\mu_{\mathrm{ren}}$ the renormalisation scale and $\gamma_E$ the Euler constant. For example for $S_{0,1,1}$ we displayed the leading order as well as the pole $1/\eps\times (\beta m)^4$, but omitted the $(\beta m)^2$-term which is irrelevant to our computation.}
\label{tab:sum-integrals}
\end{table}

\subsection{Dirac traces and contractions}
In the above, we showed how to reduce to master sum-integrals and how to compute them. We are still left with the tensorial structures to be contracted.
Let us first be more precise regarding the general form of the Dirac traces $t^{(r)}$. After symmetrisation with respect to the indices of $q$ and up to some permutations of indices and numerical factors, they will consist of terms such as
\begin{align}
\begin{split}
    t^{(0)}{}^{\mu\nu\rho\sigma} &\propto \epsilon^{\mu\nu\rho\sigma} \\
    t^{(2)}_{\alpha\beta}{}^{\mu\nu\rho\sigma} &\supset \delta^{\alpha\beta}\epsilon^{\mu\nu\rho\sigma},\;\delta^{\alpha\nu}\epsilon^{\mu\beta\rho\sigma},\, \dots \\
    t^{(4)}_{\alpha\beta\gamma\delta}{}^{\mu\nu\rho\sigma} &\supset \delta^{\alpha\beta}\delta^{\gamma\delta}\epsilon^{\mu\nu\rho\sigma},\;\delta^{\alpha\nu}\delta^{\gamma\delta}\epsilon^{\mu\beta\rho\sigma},\;\delta^{\alpha\beta}\hat\delta^{\gamma\delta}\epsilon^{\mu\nu\rho\sigma},\;\delta^{\alpha\nu}\hat\delta^{\gamma\delta}\epsilon^{\mu\beta\rho\sigma},\;\delta^{\alpha\beta}\hat\delta^{\gamma\nu}\epsilon^{\mu\delta\rho\sigma},\, \dots 
\end{split}
\end{align}
We draw attention to the presence of $\hat \delta$, which is the restriction of $\delta$ to the $d-3$ spatial dimensions which is defined in the BMHV-scheme.
We provide a non-exhaustive list of contractions needed in our computations
\begin{align}
\begin{split}
    &\delta^{\alpha\beta}\delta^0_\alpha\delta^0_\beta =1\,,\\
    &\delta^{\alpha\beta}\delta^i_\alpha\delta^j_\beta\delta_{ij}=d\,,\\
    &\hat\delta^{\alpha\beta}\delta^0_\alpha\delta^0_\beta =0\,,\\
    &\hat\delta^{\alpha\beta}\delta^i_\alpha\delta^j_\beta\delta_{ij}=d-3=-\eps\,,\\[10pt]
    &\delta^{\alpha\nu}\epsilon^{\mu\beta\rho\sigma}\delta^0_\alpha\delta^0_\beta =\delta^\nu_0\epsilon^{\mu0\rho\sigma}\,,\\
    &\delta^{\alpha\nu}\epsilon^{\mu\beta\rho\sigma} \delta^i_\alpha\delta^j_\beta\delta_{ij} =\delta^\nu_i\epsilon^{\mu i\rho\sigma}\,,\\
    &\hat\delta^{\alpha\nu}\epsilon^{\mu\beta\rho\sigma} \delta^i_\alpha\delta^j_\beta\delta_{ij} =\hat\delta^\nu_i\epsilon^{\mu i\rho\sigma} = 0 \,,\quad \text{when $\nu$ lives in dim 4}\,,\\[10pt]
    &\delta^{\alpha\beta}\delta^{\gamma\delta} \delta^i_\alpha\delta^j_\beta\delta^k_\gamma\delta^l_\delta (\delta_{ij}\delta_{kl}+\delta_{ik}\delta_{jl}+\delta_{il}\delta_{jk})=d(d+2)\,,\\
    &\delta^{\alpha\nu}\delta^{\gamma\delta} \epsilon^{\mu\beta\rho\sigma} \delta^i_\alpha\delta^j_\beta\delta^k_\gamma\delta^l_\delta (\delta_{ij}\delta_{kl}+\delta_{ik}\delta_{jl}+\delta_{il}\delta_{jk})=(d+2)\delta^\nu_i\epsilon^{\mu i\rho\sigma} \,.
    \end{split}
\end{align}

\subsection{Master sum-integral cancellations}\label{sec:cancel}
Let us now detail how the different power series in $\beta m$ of the sum-integrals end up cancelling each other in the anomalies and in parts of the currents.

There is a combination of master sum-integral that often occurs in our computation, and after massaging, it can be conveniently reduced to
\begin{align}
    \begin{split}\label{eq:idmain}
        \lim_{\eps\rightarrow 0}\Big(-(1+a \eps)S_{0,0,2}+4(1+b \eps)S_{0,1,3}+4m^2S_{0,0,3}\Big) =\frac{1}{8\pi^2}\left( 1-a+b \right) \,.
    \end{split}
\end{align}
In particular, higher powers of the denominator always cancel each other.
It is easy to prove the identity as the coefficients of the power series only slightly differ from each other
\begin{align}
    \begin{split}
        S_{0,0,2} &= \;\;\,\sum_{k=1} (\beta m)^{2k}c_k\Gamma[1/2+k] + \left(\frac{1}{8\pi^2\eps}+c\right) +\mathcal{O}(\eps) \,,\\
        S_{0,1,3} &= \;\;\,\sum_{k=1} (\beta m)^{2k}c_k\frac{1}{2}\Gamma[3/2+k] + \frac{1}{4}\left(\frac{1}{8\pi^2\eps}+\frac{1}{8\pi^2}+c\right)+\mathcal{O}(\eps) \,,\\
        m^2S_{0,0,3} &= -\sum_{k=1} (\beta m)^{2k}c_k\frac{1}{2}k\Gamma[1/2+k]+\mathcal{O}(\eps) \,,
    \end{split}
\end{align}
so that we just need to use the identity
\begin{align}
    -\Gamma[1/2+k]+2\Gamma[3/2+k]-2k\Gamma[1/2+k]=0 \,,
\end{align}
which follows from $\Gamma(z+1)=z\Gamma(z)$.

A few other similar identities are used to reduce all the results in this article and can be proved in the same way, for instance $-S_{0,0,3}+2S_{0,1,4}+2m^2S_{0,0,4} =0$.

\subsection{Axial current computational details}\label{app:jAdetail}
Here we provide intermediate results for the computation of the axial current in the massive case, which lead to the expression in terms of the function $f(\beta m)$. We recall the axial current
\begin{align}
\begin{split}
    \langle\vec j_A\rangle &=
    g_1\mu_V\vec{B}
    +g_2\vec{V}\times \vec{E}
    +g_3\mu_A\vec{B}_A
    +g_4\vec{A}\times \vec{E}_A \,,
\end{split}
\end{align}
with the coefficients that reduce to
\begin{align}
    \begin{split}
        g_1(\beta m) 
        &= \left( -S_{0,0,2} +4 S_{0,1,3}\right)=\frac{1}{2\pi^2}-f(\beta m) \,,\\[3pt]
        g_2(\beta m) 
        &= \frac{1}{d}\left( (d-4)S_{0,0,2} -4 S_{0,1,3} -4 m^2S_{0,0,3}\right)=0 \,,\\[3pt]
        g_3(\beta m) &=-\frac{2}{d (d+2)} \Big(\left(d^2-6 d+24\right) S_{0,0,2}-4\left(d^2-2 d+12\right) S_{0,1,3}-8 (d-3) S_{0,2,4}\\
        &\hspace{1cm}+\left(d^2-d+14\right) m^4 S_{0,0,4}-2 \left(d^2-5 d+26\right) m^2 \left(S_{0,0,3}-S_{0,1,4}\right)\Big)\\
        &=\frac{1}{3\pi^2}-\frac{2}{3}f(\beta m) \,,
    \end{split}
\end{align}
\begin{align}
    \begin{split}\nonumber
        g_4(\beta m) &= -\frac{2}{d (d+2)} \Big(
        +\left(d^2-10d+16\right) S_{0,0,2}+20 (d-2) S_{0,1,3}-8 (d-3) S_{0,2,4}\\
        &+\left(d^2-d+14\right) m^4 S_{0,0,4}+2 m^2 \left(\left(d^2-5 d+26\right) S_{0,1,4}-\left(d^2-7 d+22\right) S_{0,0,3}\right)\Big)\\
        &=-\frac{1}{6\pi^2}+\frac{1}{3}f(\beta m)\,.
    \end{split}
\end{align}
Once we add the pseudo-scalar mass term, the anomaly is obtained
\begin{align}
    \begin{split}
    \Acal_A[\theta]
    &=(g_1+g_2+g_{\rm{mass},1})\vec E\cdot \vec B +(g_3+g_4+g_{\rm{mass},2})\vec B_A\cdot \vec E_A \\
    &= \frac{1}{8\pi^2}(F_V \tilde F_V +\frac{1}{3} F_A \tilde F_A) \,,
    \end{split}
\end{align}
with the coefficients
\begin{align}
    \begin{split}
       (g_1+g_2+g_{\rm{mass},1})(\beta m)&= \frac{1}{d}\left( -2 (d-2) S_{0,0,2} + 4(d-1) \left(m^2 S_{0,0,3}+S_{0,1,3}\right) \right) = \frac{1}{2\pi^2} \,,\\[3pt]
        (g_3+g_4+g_{\rm{mass},2})(\beta m)&=\frac{1}{d(d+2)}( (d^2+8d-20)S_{0,0,2} +2(d^2-7d+22) S_{0,1,3} \\
        &\hspace{4.2cm}+20 m^2S_{0,0,3}+16(d-3)S_{0,2,4})\\&= \frac{1}{6\pi^2} \,.
    \end{split}
\end{align}

\bibliographystyle{JHEP}
\bibliography{biblio}
\end{document}